\documentclass[preprint,3p,12pt]{elsarticle}
\usepackage{mathrsfs}
\usepackage{amsmath}
\usepackage{stmaryrd}
\usepackage{bbding}
\usepackage{dcolumn}
\usepackage{graphicx}
\usepackage{amsfonts}
\usepackage{amssymb}
\usepackage{psfrag}
\usepackage{wrapfig}
\usepackage{subfigure}
\usepackage{makeidx}
\usepackage{bm}
\usepackage{epsf}
\usepackage{color}
\usepackage{epsfig}
\usepackage{setspace}
\usepackage{graphicx}
\usepackage{epstopdf}
\usepackage{psfrag}
\usepackage{subfigure}
\usepackage{multirow}
\usepackage{diagbox}
\usepackage{verbatim}

\usepackage{makecell}
\usepackage{float}
\usepackage{esint}
\usepackage{comment}
\usepackage{movie15}
\usepackage{hyperref}
\usepackage[ruled,linesnumbered]{algorithm2e}
\usepackage{algorithmicx}
\usepackage[noend]{algpseudocode}
\usepackage[section]{placeins}
\newtheorem{remark}{Remark}

\newcommand{\mathsym}[1]{{}}
\newcommand{\unicode}[1]{{}}
\begin{document}

\title{UGKWP and IUGKP methods for Multi-Scale Phonon Transport with Dispersion and Polarization}

	\author[HKUST1]{Hongyu Liu}
    \ead{hliudv@connect.ust.hk}
	
        \author[HKUST1]{Xiaojian Yang}
	\ead{xyangbm@connect.ust.hk}

    \author[add1]{Chuang Zhang}
\ead{zhangc520@hdu.edu.cn}

    	\author[XJTU]{Xing Ji}
	\ead{jixing@xjtu.edu.cn}
	
	\author[HKUST1,HKUST2,HKUST3]{Kun Xu\corref{cor1}}
	\ead{makxu@ust.hk}
	
	\address[HKUST1]{Department of Mathematics, Hong Kong University of Science and Technology, Clear Water Bay, Kowloon, Hong Kong}

    	\address[HKUST2]{Department of Mechanical and Aerospace Engineering, Hong Kong University of Science and Technology, Clear Water Bay, Kowloon, Hong Kong}

        	\address[HKUST3]{Shenzhen Research Institute, Hong Kong University of Science and Technology, Shenzhen, China}
	\cortext[cor1]{Corresponding author}

	\address[XJTU]{Shaanxi Key Laboratory of Environment and Control for Flight Vehicle, Xi'an Jiaotong University, Xi'an, China}

    \address[add1]{Institute of Energy, School of Sciences, Hangzhou Dianzi University, Hangzhou 310018, China}

\begin{abstract}

This paper presents two novel methods for solving multi-scale phonon transport problems with dispersion and polarization effects: the unified gas-kinetic wave-particle (UGKWP) method and the implicit unified gas-kinetic particle (IUGKP) method. Both approaches are based on solving multiple groups of BGK equations at discrete frequency points. The UGKWP method constructs multiscale macroscopic fluxes at cell interfaces through the integral solution of the unsteady BGK equation and efficiently captures non-equilibrium transport using statistical particles. Its wave-particle adaptive framework ensures computational efficiency across different regimes: in the diffusive limit, it matches the cost of explicit diffusion equation solutions, while in the ballistic limit, it performs comparably to pure particle methods. The IUGKP method, specifically designed for steady-state problems, determines particle evolution scale based on the physical mean free path. This approach enables rapid convergence at both large and small Knudsen numbers, with the latter facilitated by a newly constructed macroscopic prediction equation. Both methods incorporate an adaptive frequency-space sampling technique that maintains particle counts per cell comparable to single-frequency methods, significantly improving computational efficiency and memory usage. The accuracy and efficiency of both methods are validated through various numerical tests, including large-scale three-dimensional conduction heat transfer simulations. Results demonstrate their effectiveness in handling complex phonon transport phenomena across multiple scales.

\end{abstract}


\maketitle

\section{Introduction}
Due to the significant challenges posed by high integration and power densities in micro- and nano-scale electronic devices, effective thermal management becomes exceptionally crucial~\cite{TCAD_application_intel_2021_review,murthy2005review,review_2019_CMOS,HUA2023chapter}.
However, at the micro- and nano-scale, Fourier's law of heat conduction fails, necessitating more fundamental equations to accurately describe heat transfer~\cite{phononsnanoscale,RevModPhys.90.041002,RevModPhys.94.025002}.
The phonon Boltzmann transport equation (BTE) is a more fundamental equation that accurately captures the free transport and scattering processes of phonons.
Consequently, it serves as a powerful tool for describing heat conduction at the micro- and nano-scale~\cite{murthy2005review, ziman2001electrons, chattopadhyay2014comparative}.
In many cases, particularly for high-dimensional problems, obtaining an analytical solution to the phonon BTE is very challenging. As a result, numerical methods have become the preferred approach for solving it.
The BTE equation have six dimensions in total: three spatial dimensions, two angular dimensions, and one frequency dimension, which means that solving the equation often requires huge computational time and memory~\cite{barry2022boltzmann, mazumder_boltzmann_2022, guo_progress_DUGKS, peraud_monte_2014, lacroix2005monte}.

Various numerical methods have been developed to solve the phonon BTE, generally classified as deterministic or stochastic statistical methods.
Deterministic approaches include methods such as the lattice Boltzmann method \cite{GuoZl13LB}, discrete ordinate method (DOM) \cite{stamnes1988dom, murthy1998domunstruct, SyedAA14LargeScale, FivelandVA96Acceleration}, (discrete) unified gas kinetic scheme (UGKS/DUGKS) \cite{guo_progress_DUGKS, guo2016dugksphonon, luo2017dugksphonon, zhang2019dugksphonon}, synthetic iterative scheme \cite{Chuang17gray, zhang2023acceleration, zhang2025synthetic}, among others.
The LBM method is based on a near-equilibrium assumption \cite{qian1992lattice}, which makes it suitable for thermal conduction problems at low Knudsen numbers; however, its computations are not accurate for high Knudsen number regimes \cite{GuoZl13LB, chattopadhyay2014comparative}.
The DOM method decouples the scattering/collision and streaming processes of the evolution process of distribution function, which necessitates that the time step should be smaller than the relaxation time. Consequently, as the Knudsen number decreases, the computational cost of the DOM method becomes prohibitively high.
The UGKS and DUGKS methods, based on the integral solution of the BGK equation, couple the transport and collision processes of the distribution function. Their multi-scale flux enables the grid size and time step to be free from the restrictions of the mean free path and relaxation time.
But, all the deterministic methods mentioned above often suffer from the curse of dimensionality, encountering memory bottlenecks in high-dimensional BTE problems and making large-scale simulations challenging.
On the other hand, stochastic statistical methods are exemplified by the Monte Carlo \cite{DSMC_book_1994,DSMC_phonon_1994, mazumder2001monte, lacroix2005monte,mittal2010monte, PJP11MC, peraud2015derivationmonte, peraud_monte_2014} method.
The MC method uses statistical particles to replace the integration over a high-dimensional discretized angular space, significantly reducing memory usage and making it more efficient than the explicit DOM method in high Knudsen number regimes. However, since the MC method also decouples collisions from streaming just like the explicit DOM, the computational time step or grid size must be smaller than the relaxation time or mean free path due to the numerical accuracy. As the Knudsen number gradually decreases, the computational cost of the MC method becomes prohibitively high due to frequent collisions\cite{peraud_monte_2014, lacroix2005monte, PJP11MC, SILVA2024108954, PATHAK2021108003}.
In addition, it suffers serious statistical errors.

As the wave-particle version of the UGKS method, the UGKWP method employs a wave-particle adaptive evolution framework that combines the efficiency of macroscopic solvers with the particle approach's ability to capture non-equilibrium effects, saving memory and enhancing computational speed. The UGKWP method has been widely applied to multiscale neutral gas transport \cite{zhu2019ugkwp}, plasmas \cite{liu2021ugkwp, pu2025ugkwp}, multiphase flows \cite{yang2024solid}, radiation \cite{yang2025rad}, and turbulence simulation \cite{yang2025turb}.
Moreover, our previous work has extended the UGKWP method to solve the phonon transport equation for the gray model~\cite{liu2025unified}.
Please note that the wave-particle approach discussed here differs from that in quantum mechanics~\cite{phononsnanoscale,kaviany_2008}. The phonon BTE describes only the particle aspect of phonon wave-particle duality, while the ``wave''  refers to the macroscopic temperature wave phenomenon.

Building on our previous work, this paper further develops the UGKWP method to solve the phonon transport equation, taking into account phonon dispersion relations and polarizations.
The UGKWP method proposed in this paper solves the multiple groups of BGK equations at discrete frequency points.
In each frequency point, it similar as solving a gray model BGK equation.
However, due to differences in the physical models, the frequency-dependent phonon transport model and the gray phonon transport model exhibit distinct definitions for the equilibrium state, as well as different collision compatibility conditions.
In fact, each frequency is associated with its own Knudsen number.
The UGKWP method’s wave-particle adaptive evolution framework allows it to solve the equations corresponding to different Knudsen number regimes for each frequency.
When a particular frequency has a large Knudsen number, the UGKWP method essentially operates as a pure particle-based method.
Conversely, when the Knudsen number at a certain frequency approaches zero, the flux automatically reverts to typical Fourier’s law, meaning that there are no particles in UGKWP.
However, for steady-state problems, the UGKWP method is constrained by the physical time step.
To address this limitation, this work further develops the IUGKP method for solving steady-state problems.
In the IUGKP method, the scale of particle evolution is determined by the physical mean free path.
As a result, the method converges fast when the Knudsen number is large.
Conversely, when the Knudsen number is small, the macroscopic prediction equation constructed in this paper plays a dominant role, also ensuring fast convergence.
Additionally, to further reduce the number of particles, this work introduces an adaptive sampling method in the frequency space.
This method allows the number of particles per cell in the IUGKP method and the UGKWP method to remain within the same order of magnitude as that used in single-frequency methods. Consequently, it significantly accelerates the computational speed and also saves computing memory.

This paper is organized as follows.
Section 2 introduces the frequency dependent phonon BGK equation.
Section 3 presents the UGKWP method for solving the frequency-dependent phonon BGK equation.
Section 4 introduces the IUGKP method for solving the steady-state frequency-dependent phonon BGK equation.
Section 5 is the numerical examples.
The last section is the conclusion.

\section{Phonon Boltzmann transport equation}
The phonon BTE is an effective tool for describing the heat conduction in solid materials from the ballistic to diffusive regimes.
This section will first introduce the frequency-dependent phonon Boltzmann equation (BTE) and then simplify it into a form based on energy derivation.

 In general, the equation can be simplified using the BGK type relaxation time approximation model \cite{BGK, ziman2001electrons, Chuang17gray, PJP11MC}
\begin{equation}
\frac{\partial f}{\partial t}+\boldsymbol{V_g} \cdot \nabla f=\frac{1}{\tau}\left(f^{e q}-f\right),
\label{eq:BGKBTE}
\end{equation}
where $f$ is the phonon distribution function, $\boldsymbol{V_g}=|\boldsymbol{V_g}| \boldsymbol{s}$ is the group velocity, $\boldsymbol{s}= (\cos \theta, \\ \sin \theta \cos \varphi, \sin \theta \sin \varphi)$ is the unit directional vector $(\theta$ is the polar angle and $\varphi$ is the azimuthal angle), $\tau$ is the relaxation time, $f^{eq}$ is the equilibrium distribution function, which satisfy the Bose-Einstein distribution,
\begin{equation}\label{bose-einstein}
f^{eq} (T)=\frac{1}{\exp \left(\hbar \omega / k_B T\right)-1},
\end{equation}
where $\hbar$ is the reduced Planck constant, $\omega$ is the frequency, $k_B$ is the Boltzmann constant and $T$ is the temperature.

For convenience, we define a new distribution function as:
\begin{equation}
e(\boldsymbol{x}, \boldsymbol{s}, \omega, p)=\hbar \omega\left(f-f^{e q}\left(T_{\mathrm{ref}}\right)\right) D(\omega, p) / 4 \pi,
\end{equation}
where $D(\omega, p)$ is the phonon state density, and it satisfies:
\begin{equation}
    D(\omega, p)=k^2 /\left(2 \pi^2|\boldsymbol{V_g}|\right).
\end{equation}
This new distribution function represents the energy difference from the equilibrium state at temperature $T_{ref}$. It assumes that, for the same phonon branch and frequency, the phonon particle velocities are uniformly distributed on the spherical surface.
Then, the corresponding equilibrium state is:
\begin{equation}
e^{e q}(\boldsymbol{x}, \omega, p)=\hbar \omega\left(f^{e q}-f^{e q}\left(T_{\mathrm{ref}}\right)\right) D(\omega, p) / 4 \pi.
\end{equation}
At the same time, by performing a Taylor expansion of the equilibrium state at the $T_{ref}$ point, we can obtain:
\begin{equation}\label{cv-from-taylor}
    f^{e q}\left(T\right)-f^{e q}\left(T_{\mathrm{ref}}\right) = 
    \frac{\partial f^{e q}} {\partial T}|_{T_{\mathrm{ref}}}\left(T-T_{\mathrm{ref}}\right).
\end{equation}
Thus, the new equilibrium distribution function can be written as:
\begin{equation}
e^{e q}(\boldsymbol{x}, \omega, p)=C_g(\boldsymbol{x}, \omega, p)(T-T_{\mathrm{ref}}) / 4 \pi,
\end{equation}
where $C_g(\boldsymbol{x}, \omega, p)$ is the specific heat at branch $p$ and frequency $\omega$, and it can be written as:
\begin{equation}
C_g(\boldsymbol{x}, \omega, p)=\hbar \omega\frac{\partial f^{e q}} {\partial T}|_{T_{\mathrm{ref}}}D(\omega, p).
\end{equation}
Then the simplified frequency-dependent phonon BGK can be written as:
\begin{equation}\label{energy-bgk}
\frac{\partial e_g}{\partial t}+\boldsymbol{V_g} \cdot \nabla e_g=\frac{1}{\tau_g}\left(e^{e q}_g-e_g\right),
\end{equation}
where the subscripts indicate the phonon branch $p$ and the frequency $\omega$.

The energy $E_g$ for phonon branch $p$ and frequency $\omega$ can be obtained by integrating the whole angular space:
\begin{equation}\label{Eg-integral}
    E_g=\iint_{4\pi} e_g \mathrm{~d} \Omega.
\end{equation}
It is worth noting that, for the frequency-dependent phonon Boltzmann equation, the compatibility condition is different from that in the gray model. 
For the collision capability condition, we have:
\begin{equation}
\sum_p \int_{\omega_{\min , p}}^{\omega_{\max , p}} \iint_{4 \pi} \frac{e^{e q}_g-e_g}{\tau_g} \mathrm{~d} \Omega\mathrm{~d}\omega=0.
\end{equation}
From the compatibility condition, we have:
\begin{equation}
\begin{aligned}
T= & T_{\mathrm{ref}}+\left(\sum_p \int_{\omega_{\min , p}}^{\omega_{\max , p}} \frac{\iint_{4 \pi} e_g \mathrm{~d} \Omega}{\tau_g} \mathrm{~d} \omega\right) \\
& \times\left(\sum_p \int_{\omega_{\min , p}}^{\omega_{\max , p}} \frac{C_g}{\tau_g} \mathrm{~d} \omega\right)^{-1}.
\end{aligned}
\end{equation}
From this equation, each time we update $E_g$, the new temperature can be obtained by this equation, and the new equilibrium distribution function can be obtained.

Moreover, the total energy is an integral part of the solid angle and frequency space:
\begin{equation}
\boldsymbol{E}=\sum_p \int_{\omega_{\text {min }, p}}^{\omega_{\text {max } p}} \iint_{4 \pi} e_g\mathrm{~d} \Omega \mathrm{~d} \omega.
\end{equation}
Meanwhile, the total heat flux is:
\begin{equation}
\boldsymbol{q}=\sum_p \int_{\omega_{\text {min }, p}}^{\omega_{\text {max } p}} \iint_{4 \pi} \boldsymbol{V_g} e_g \mathrm{~d} \Omega \mathrm{~d} \omega.
\end{equation}

In practical numerical calculations, it is challenging to compute integrals in frequency space analytically. Therefore, we discretize the frequency space and replace the integral with numerical integration.

Thus, the total energy and heat flux can be obtained by:
\begin{equation}
\boldsymbol{E}=\sum_p \sum_{\omega} \iint_{4 \pi} e_g\mathrm{~d} \Omega \Delta\omega_g,
\end{equation}
\begin{equation}
\boldsymbol{q}=\sum_p \sum_{\omega} \iint_{4 \pi} \boldsymbol{V_g} e_g \mathrm{~d} \Omega \Delta\omega_g,
\end{equation}
where $\omega_g$ is the interval length of the $g^{th}$ frequency domain.

Moreover, the temperature can be obtained by:
\begin{equation}\label{temperature-update}
\begin{aligned}
T= & T_{\mathrm{ref}}+\left(\sum_p \sum_{\omega} \frac{\iint_{4 \pi} e_g \mathrm{~d} \Omega}{\tau_g} \Delta\omega_g\right) \\
& \times\left(\sum_p \sum_{\omega} \frac{C_g}{\tau_g} \Delta\omega_g\right)^{-1}.
\end{aligned}
\end{equation}

\section{Multi-scale UGKWP method for frequency-dependent phonon transport}
In this section, we first introduce the time-dependent multi-scale UGKWP algorithm for frequency-dependent phonon transport.

In the UGKWP method, the cell-averaged values are evolved using a finite volume approach. By performing an angular integration on the $g^{th}$ group phonon BTE and then applying finite volume spatial discretization, the energy evolution equation can be obtained:
\begin{equation}
     E_{g,i}^{n+1}-E_{g,i}^{n}+\frac{1}{V_i}\sum_{j \in N(i)} S_{i j} \boldsymbol{F}_{g,i j}=\int_0^{\Delta t}\frac{\iint_{4\pi} e^{eq}_g-e_g \mathrm{~d}\Omega}{\tau_g} \mathrm{~d} t,
\end{equation}
where $V_i$ is the volume of cell $i$,  $S_{ij}$ is the area of the $j$-th interface of cell $i$. $E_{g,i}$ is the associated $g^{th}$ group cell-averaged energy. $F_{g,i j}$ denotes the $g^{th}$ group macroscopic flux across the interface $S_{i j}$, which can be written as
\begin{equation}
    \boldsymbol{F}_{g,i j}=\int_0^{\Delta t}\iint_{4\pi} \boldsymbol{V_g} \cdot \boldsymbol{n}_{ij} e_{g,ij} \mathrm{~d} \Omega \mathrm{~d} t,
\end{equation}
where $\boldsymbol{n}_{ij}$ is the unit normal vector pointing out of the cell.
It is important to note that, unlike the gray model. Although the equation for each group can be regarded as a gray phonon BGK equation, the evolution of each group does not satisfy the collision compatibility conditions. Instead, it is the cumulative effect of all groups that satisfies the collision compatibility conditions, as mentioned above.

To construct the multiscale interface flux, we need to obtain the correct distribution function at the cell interface. 
Based on the integral solution, the interface distribution function can be written as:
\begin{equation}\label{integral-solution-unsteady}
    e^{eq}_{g}(\mathbf{0}, t, \boldsymbol{V_g})=\frac{1}{\tau_{g}} \int_0^t e^{e q}_{g}\left(x^{\prime}, t^{\prime}, \boldsymbol{V_g}\right) e^{-\frac{t-t^{\prime}}{\tau_{g}}} \mathrm{d} t^{\prime}+e^{-\frac{t}{\tau_{g}}} e_0(-\boldsymbol{V_{g}} t, \boldsymbol{V_{g}}),
\end{equation}
where "0" means the cell interface.

This integral solution describes the transport process of particles arriving at the terminal point along characteristic lines. Specifically, particles with their initial distribution have a probability $e^{-\frac{t}{\tau}}$ of preserving that distribution while moving toward the terminal point. If a collision occurs at an intermediate time $t^{\prime}$, the particles have a probability $\frac{1}{\tau}e^{-\frac{t-t^{\prime}}{\tau}}$ of moving to the interface according to the local equilibrium distribution.
This process fundamentally differs from the DOM method. The interface distribution function, obtained through the integral solution, couples the particle migration and collision processes. Consequently, the method is not constrained by the mean free path, and the constructed interface flux is multiscale.
Furthermore, performing a first-order Taylor expansion in time and space along the characteristic line for $e^{e q}_g(\boldsymbol{x}, t)$ as Eq.~\eqref{taylor-equlibrium}:

\begin{equation} \label{taylor-equlibrium}
    e^{e q}_g(\boldsymbol{x}, t)=e_{g,0}^{e q}+e_{g,t}^{e q} t+e_{g,x}^{e q} \cdot \boldsymbol{V_g}t.
\end{equation}
Substituting it into Eq.~\eqref{integral-solution-unsteady}, the explicit formula for distribution function at time $t$ can be obtained as Eq.~\eqref{time-dependent-distribution-function}
\begin{equation} \label{time-dependent-distribution-function}
e_g(t)=c_{g,1} e_{g,0}^{e q}+c_{g,2} e_{g,x}^{e q} \cdot \boldsymbol{V_g}+c_{g,3} e_{g,t}^{e q}+c_{g,4} e_{g,0}+c_{g,5} e_{g,\boldsymbol{x}} \cdot \boldsymbol{V_g}.
\end{equation}
Taylor expansion in space for the initial distribution function is also considered in Eq.~\eqref {time-dependent-distribution-function} with the coefficients
\begin{equation}
    \begin{aligned}
& c_{g,1}=1-e^{-t / \tau_g} \\
& c_{g,2}=t e^{-t / \tau_g}-\tau\left(1-e^{-t / \tau_g}\right) \\
& c_{g,3}=t-\tau_g\left(1-e^{-t / \tau_g}\right) \\
& c_{g,4}=e^{-t / \tau_g} \\
& c_{g,5}=-t e^{-t / \tau_g}.
\end{aligned}
\end{equation}

Then, the multi-scale macro numerical flux passing the interface can be written as:
\begin{equation}
\begin{aligned}
\int_0^{\Delta t} \iint_{4\pi} \boldsymbol{V_g} \cdot \boldsymbol{n}_{i j} e_{g,i j}(t)  \mathrm{~d} \Omega \mathrm{~d} t= & \iint_{4\pi} \boldsymbol{V_g} \cdot \boldsymbol{n}_{i j}\left(q_{g,1} e_{g,0}^{e q}+q_{g,2} e_{g,\boldsymbol{x}}^{e q} \cdot \boldsymbol{V_g}+q_{g,3} e_{g,t}^{e q}\right) \\
& +\boldsymbol{V_g} \cdot \boldsymbol{n}_{i j}\left(q_{g,4} e_{g,0}+q_{g,5} e_{g,\boldsymbol{x}} \cdot \boldsymbol{V_g}\right) \mathrm{~d} \Omega  \\
= & \mathcal{F}_{g,i j}^{e q}+\mathcal{F}_{g,i j}^{f r},
\end{aligned}
\end{equation} with the coefficients
\begin{equation}
    \begin{aligned}
& q_{g,1}=\Delta t-\tau_g\left(1-e^{-\Delta t / \tau_g}\right), \\
& q_{g,2}=2 \tau_g^2\left(1-e^{-\Delta t / \tau_g}\right)-\tau_g \Delta t-\tau_g \Delta t e^{-\Delta t / \tau_g}, \\
& q_{g,3}=\frac{\Delta t^2}{2}-\tau_g \Delta t+\tau_g^2\left(1-e^{-\Delta t / \tau_g}\right), \\
& q_{g,4}=\tau_g\left(1-e^{-\Delta t / \tau_g}\right), \\
& q_{g,5}=\tau_g \Delta t e^{-\Delta t / \tau_g}-\tau_g^2\left(1-e^{-\Delta t / \tau_g}\right) .
\end{aligned}
\end{equation}

Moreover, $|\boldsymbol{V}_g|$ in phonon transport of a certain group $g$ is a constant. 
Thus, $\boldsymbol{V_g}=|\boldsymbol{V}_g| s, s=(\cos \theta, \sin \theta \cos \varphi, \sin \theta \sin \varphi)$. 
Meanwhile, we can get the micro derivative of $e_g^{e q}$ using the chain rule:
\begin{equation}\label{micro-slope}
    \frac{\partial e_g^{e q}}{\partial x}=\frac{\partial e_g^{e q}}{\partial T} \frac{\partial T}{\partial x}=\frac{1}{4 \pi} \frac{C_{gv}\partial T}{\partial x} .
\end{equation}

Similar to the gray model, the flux of equilibrium for the current phonon group only retains the spatial derivative term.
The other two terms vanished due to the symmetry properties.
The specific formula of $\mathcal{F}_{g,i j}^{e q}$ is:
\begin{equation}\label{prim-Feq}
\mathcal{F}_{g,i j }^{e q}=\iint_{4\pi} \boldsymbol{V_g} \cdot \boldsymbol{n}_{i j}\left(q_{g,2} e_{g,\boldsymbol{x}}^{e q} \cdot \boldsymbol{V_g}\right) \mathrm{~d} \Omega,
\end{equation}
and substituting Eq~.\eqref{micro-slope} into Eq~.\eqref{prim-Feq}, $\mathcal{F}_{g,i j}^{e q}$ can be written as:
\begin{equation}
    \mathcal{F}_{g,i j}^{e q}= \frac{q_{g,2}}{4 \pi} \iint_{4\pi} \boldsymbol{V_g} \cdot \boldsymbol{n}_{i j}\left(C_{g,v}\nabla_x T \cdot \boldsymbol{V_g}\right) \mathrm{~d} \Omega.
\end{equation}

It can be further simplified as:
\begin{equation}
\mathcal{F}_{g,i j}^{e q}=\frac{C_{g,v}q_{g,2} |\boldsymbol{V_g}|^2}{4 \pi} \iint_{4\pi} s \cdot n_{i j}\left(\nabla_x T \cdot s\right) \mathrm{~d} \Omega
\end{equation}

Then, integrating it on the whole angular space, $\mathcal{F}_{g,i j}^{e q}$ can be written as:
\begin{equation}
\mathcal{F}_{g,i j}^{e q}=\frac{C_{g,v}q_{g,2} |\boldsymbol{V_g}|^2}{3} \boldsymbol{n}_{i j} \cdot \nabla_x T
\end{equation}

Recall the compatibility condition used in GKS (Gas Kinetic Scheme):
\begin{equation}\label{capability-condition}
    \iint_{4\pi} e_g^{eq} \mathrm{~d} \Omega = \iint_{\boldsymbol{V_n}>0} e_g^{eq,L} \mathrm{~d} \Omega +\iint_{\boldsymbol{V_n}<0} e_g^{eq,R} \mathrm{~d} \Omega,
\end{equation}
where $\boldsymbol{V}_n$ is the normal component of the phonon group velocity at the cell interface. 

Then take spatial derivative of Eq.~\eqref{capability-condition} on both side:
\begin{equation}\label{capability-condition-derivative}
e_{g,x}^{e q}=\frac{1}{4 \pi}\left(\iint_{\boldsymbol{V_n}>0} e_{g,x}^{e q,L}  \mathrm{~d} \Omega+\iint_{\boldsymbol{V_n}<0} e_{g,x}^{e q,R}  \mathrm{~d} \Omega\right),
\end{equation}
which can be further simplified as:
\begin{equation}\label{half-micro-slope}
e_{g,x}^{e q}=\frac{e_{g,x}^{e q,L}+e_{g,x}^{e q,R}}{2}.
\end{equation}

Substituting Eq~.\eqref{micro-slope} into Eq~.\eqref{half-micro-slope}, the slope of temperature can be obtained:

\begin{equation}
\left(\frac{\partial T}{\partial x}\right)^0=\frac{\left(\frac{\partial T}{\partial x}\right)^L+\left(\frac{\partial T}{\partial x}\right)^R}{2}
\end{equation}

Finally, the equilibrium transport flux can be written as:
\begin{equation}
    \mathcal{F}_{g,i j}^{e q}=\frac{C_{g,v}q_{g,2} |\boldsymbol{V_g}|^2}{6} \left[\left(T_n\right)_L+\left(T_n\right)_R\right],
\end{equation}
where $\left(T_n\right)_L$ and $\left(T_n\right)_R$ are the left and right point values of normal temperature gradients. 

In non-equilibrium transport, free transport along the characteristic line generates the free transport flux. 
UGKS stores distribution functions in each cell and uses spatial reconstruction to obtain the left and right values of the distribution function at the interface to compute the flux, and UGKP employs statistical particles whose propagation along the characteristic line forms the flux. The particle trajectory is described as Eq.~\eqref{particle-path}

\begin{equation}\label{particle-path}
\boldsymbol{x}=\boldsymbol{x}^n+\boldsymbol{V_g} t_f,
\end{equation}
where $t_f$ denotes the phonon's free streaming time.
Recall that the integral solution indicates that a particle has a probability $e^{-\frac{t}{\tau}}$ of maintaining its initial state as it propagates from the initial time to time t.
Thus, the particle's free transport time can be sampled as follows:

\begin{equation}
t_f=\min \left[-\tau_g \ln (\eta), \Delta t\right] ,
\end{equation}
where $\eta$ represents a random number uniformly distributed over the interval $[0, 1]$.

Therefore, the evolution of the macroscopic energy for $g^{th}$ group can be described by the following equation:
\begin{equation}
E_{g,i}^{n+1}-E_{g,i}^n=-\frac{1}{V_i} \sum_{j \in N(i)} S_{i j} \mathcal{F}_{g,i j}^{e q} + \frac{1}{V_i}\left(\sum w_{g,p}^{in} - \sum w_{g,p}^{out}\right) + \int_0^{\Delta t}\frac{\iint_{4\pi} e^{eq}_g-e_g \mathrm{~d}\Omega}{\tau_g} \mathrm{~d} t.
\end{equation}
where $w_{g,p}^{in}$ means the energy of the particle entering the target cell and $w_{g,p}^{out}$ means the energy of the particle leaving the target cell for the $g^{th}$ group.

Based on the free transport time sampled for the particles, they can be divided into collisionless and collisional particles.
Collisionless particles refer to those that are transported without collisions during the physical time step $\Delta t$, meaning their free transport time is exactly $\Delta t$. In contrast, within the same physical time step $\Delta t$, collisional particles first undergo free transport for a duration $t_f$ and then experience a series of collisions.
Thus, the collisional particles are removed at the transport endpoint upon completing free transport, and their corresponding energy is counted into the relevant cell's total macroscopic energy.
From the updated cell total energy, the unsampled  particles are:
\begin{equation}
    E^h_g = E^{n+1}_g -\frac{1}{V_i} \sum w^{fr}_g,
\end{equation}
where $E^h_g$ is the summation of the umsampled particles' energy of $g^{th}$ group, $w^{fr}_g$ is the energy of the collisionless particle of $g^{th}$ group.
Moreover, $E^h_g$ can be recovered by the particle re-sampling process as Eq.~\eqref{kp-sample} illustrates:

\begin{equation}\label{kp-sample}
    w^p_g=\frac{E^h_gV_i}{N},
\end{equation}
where N is the re-sample number, $w^p_g$ is the particle's energy of $g^{th}$ group.

The method described above is the UGKP method, in which, as previously stated, all the free-transport fluxes are entirely represented by statistical particles.

The UGKWP method extends UGKP by more efficiently representing non-equilibrium transport. 
Instead of modeling all non-equilibrium transport with particles, UGKWP decomposes it into wave and particle components, further reducing the number of sampled particles.

In UGKP, the resampled particles' free transport flux can be analytically written as:
\begin{equation}
\begin{aligned}
    F_{g,i j}^{f r, U G K S}\left(e^{eq,h}_g\right)&=\int_0^{\Delta t}\iint_{4\pi} \boldsymbol{V_g} \cdot \boldsymbol{n}_{ij} e^{-\frac{t}{\tau_g}} e^{eq,h}_g\left(-\boldsymbol{V_g}t,\boldsymbol{V_g}\right) \mathrm{~d} \Omega \mathrm{~d} t \\
    &= \iint_{4\pi} \boldsymbol{V_g} \cdot \boldsymbol{n}_{ij}  \left[ q_{g,4} e^{eq,h}_g\left(0,\boldsymbol{V_g}\right) + q_{g,5} \boldsymbol{V_g} \cdot e_{g,x}^{eq,h}\left(0,\boldsymbol{V_g}\right)\right]\mathrm{~d} \Omega.
\end{aligned}
\end{equation}

Meanwhile, the UGKWP method only resamples collisionless particles to maintain non-equilibrium transport.
Thus, the energy for collisionless particles is:
\begin{equation}
    e^{hp}_g=e^{-\frac{\Delta t}{\tau_g}}E^h_g.
\end{equation}
In a statistical sense, the free transport of these collisionless particles is equivalent to the flux contributed by the distribution function carrying the corresponding energy in the DOM (Discrete Ordinates Method) framework, as Eq.~\eqref{fr-dvm} shows:
\begin{equation}\label{fr-dvm}
\begin{aligned}
    F_{g,i j}^{f r, D O M }\left(e^{eq,hp}_g\right)&=e^{-\frac{\Delta t}{\tau_g}}\int_0^{\Delta t}\iint_{4\pi} \boldsymbol{V_g} \cdot \boldsymbol{n}_{ij}  e^{eq,h}_g\left(-\boldsymbol{V_g}t,\boldsymbol{V_g}\right) \mathrm{~d} \Omega \mathrm{~d} t \\
    &= e^{-\frac{\Delta t}{\tau_g}}\iint_{4\pi} \boldsymbol{V_g} \cdot \boldsymbol{n}_{ij}  \left[  \Delta t e^{eq,h}_g\left(0,\boldsymbol{V_g}\right) -\frac12 \Delta t^2 \boldsymbol{V_g} \cdot  e_{g,x}^{eq,h}\left(0,\boldsymbol{V_g}\right)\right]\mathrm{~d} \Omega.
\end{aligned}
\end{equation}

Therefore, the portion of the non-equilibrium flux arising from the wave component, excluding the free transport contribution of collisionless particles, is given by:
\begin{equation}
\begin{aligned}
F_{g,i j}^{f r, w a v e} & =F_{g,i j}^{f r, U G K S}\left(e^{eq,h}_g\right)-F_{i j}^{f r, D O M}\left(e^{eq,h p}_g\right) \\
& =\iint_{4\pi} \boldsymbol{V_g} \cdot \boldsymbol{n}_{i j}\left[\left(q_{g,4}-\Delta te^{-\frac{\Delta t}{\tau_g}}\right) e^{eq,h}_g(\mathbf{0}, \boldsymbol{V_g})
+\left(q_{g,5}+\frac{\Delta t^2}{2} e^{-\frac{\Delta t}{\tau_g}}\right) \boldsymbol{V_g} \cdot e_{g,x}^{eq,h}(\mathbf{0}, \boldsymbol{V_g})\right] \mathrm{~d} \Omega.
\end{aligned}
\end{equation}

Moreover, $F_{g,i j}^{f r, w a v e}$ can be calculated explicitly as:
\begin{equation}\label{fr,wave}
\begin{aligned}
        F_{g,i j}^{f r, w a v e}&=\left(q_{g,4}-\Delta t e^{-\Delta t / \tau_g}\right) \frac{\boldsymbol{V_g}\left(E_{g,L}^h-E_{g,R}^h\right)}{4} \\
    &+ \left(q_{g,5}+\frac{\Delta t^2}{2} e^{-\Delta t / \tau_g}\right)\left[\frac{\boldsymbol{V_g}^2}{6}\left(\frac{\partial E^h_g}{\partial n}\right)^R+\frac{c^2}{6}\left(\frac{\partial E^h_g}{\partial n}\right)^L\right].
\end{aligned}
\end{equation}

Finally, the evolution of the UGKWP method for cell i and $g^{th}$ group can be written as:

\begin{equation}
\begin{aligned}
    E_{g,i}^{n+1}-E_{g,i}^n&=-\frac{1}{V_i} \sum_{j \in N(i)} S_{i j} \left(\mathcal{F}_{g,i j}^{e q}+\mathcal{F}_{g,i j}^{fr,wave}\right)+\frac{1}{V_i}\left(\sum w_{g,p}^{i n}-\sum w_{g,p}^{out}\right)\\ 
    &+ \int_0^{\Delta t}\frac{\iint_{4\pi} e^{eq}_g-e_g \mathrm{~d}\Omega}{\tau_g} \mathrm{~d} t.
\end{aligned}
\end{equation}

Further description of particle re-sampling and deletion processes is necessary to provide a more comprehensive introduction to the UGKWP method.
In this study, a reference number method \cite{zhu2019ugkwp} is employed for particle sampling to control the number of sampled particles.
Given the reference number, the reference sampling energy for a particle can be written as:
\begin{equation}
w_{g,r}=\frac{\left(E_g-E^h_g\right)+E^h_g e^{-\Delta t / \tau_g}}{N_r} V,
\end{equation}
where $w_{g,r}$ is the reference sampling energy for a particle of $g^{th}$ group, $N_r$ is the reference number.

As illustrated above, the exact re-sampling energy from the wave is $E^h_ge^{-\frac{\Delta t}{\tau_g}}$. So, the exact sampling number is:
\begin{equation}
N_{g,s}= \begin{cases}0, & \text { if } V e^{-\frac{\Delta t}{\tau_g}} E^h_g \leq e_{g,\min } \\ 2\left\lceil\frac{e^{-\frac{\Delta t}{\tau_g}} E^h_g}{2\left(E_g-E^h_g\right)+2 E^h_g e^{-\Delta t / \tau_g}} N_{g,ref}\right\rceil, & \text { if } V e^{-\frac{\Delta t}{\tau_g}} E^h_g>e_{g,\min }\end{cases},
\end{equation}
As a result, the energy of each particle is $\frac{e^{-\Delta t / \tau_g} E^h_g V}{N_{g,s}}$ and $e_{min}=0.0001E_g$.
Additionally, the velocity directions of these particles are uniformly distributed over the spherical domain.
This can be obtained by the Inverse Sampling method:
\begin{equation}
\theta=\arccos \left(1-2 \eta_1\right),
\end{equation}
\begin{equation}
\varphi=2 \pi \eta_2,
\end{equation}
where $\eta_1$ and $\eta_2$ are random numbers uniformly distributed in $[0,1]$.

For clarity and reproducibility, Algorithm~.\ref{wp-algorithm} presents a summary of the UGKWP method for frequency-dependent phonon transport:
\begin{algorithm}\label{wp-algorithm}
	\caption{UGKWP method for frequency-dependent phonon transport}
	\label{wp-algorithm}
	Initialization: set the initial temperature and energy field for each group\;
	  \textbf{while} ($t<t_{stop}$) \textbf{do}\;
        classify the remaining particles by compare $t_f$ and $\Delta t$\;
        Sample free stream particles both for the inner region and boundary\;
        spatial reconstruction for $T$ and $E^h_g$\;
        Calculate macroscopic flux $\mathcal{F}_{g,i j}^{e q}$ and $\mathcal{F}_{g,i j}^{f r, w a v e}$\;
        Particle free streaming\;
        Update Macroscopic energy by evolution equation\;
        Update temperature by Eq~.\eqref{temperature-update}\;
        Delete collisional particles which $t_f<\Delta t$\;
        \textbf{end while}\;
        Output
\end{algorithm}	

\section{IUGKP method for frequency-dependent phonon transport}
The previously introduced UGKWP method for frequency-dependent phonon transport in steady-state problems suffers from convergence limitations due to the time step size.
In this section, we develop a corresponding multiscale acceleration algorithm, IUGKP, aiming to solve the steady-state frequency-dependent phonon transport problem more efficiently.
In our previous work, we introduced the IUGKP method for steady-state gray phonon transport, achieving high acceleration across various scales. Building on these ideas, this section extends the method to frequency-dependent phonon transport systems.

\subsection{Steady state phonon BGK equation}
The simplified energy derivation based steady phonon BGK equation is:
\begin{equation}
\boldsymbol{V}_g \cdot \nabla e_g=\frac{1}{\tau_g}\left(e_g^{e q}-e_g\right).
\end{equation}

Different from the unsteady BGK equation's integral solution, the integral solution for steady state BGK equation can be written as:
\begin{equation}
e_g(x, u_g)=\int_{x_0}^x \frac{e_g^{eq}}{u_g \tau_g\left(x^{\prime}\right)} \mathrm{e}^{-\int_{x^{\prime}}^x \frac{1}{u_g \tau_g\left(x^{\prime \prime}\right)} \mathrm{d} x^{\prime \prime}} \mathrm{d} x^{\prime}+e_g\left(x_0, u_g\right) \mathrm{e}^{-\int_{x_0}^x \frac{1}{u_g \tau_g\left(x^{\prime}\right)} \mathrm{d} x^{\prime}},
\end{equation}
where $x_0$ is the starting point and the relaxation time $\tau_g$ is related to space location $x$, $u_g$ is the magnitude of the group velocity.

The integral solution of the $g^{th}$ group steady BGK equation demonstrates that the particles located at the starting point have a cumulative density function (CDF) $P_{g,f}$ to maintain their distribution while free-streaming to the endpoint $x$; the formulation of $P_{g.f}\left(x_0,x\right)$ is:
\begin{equation}
    P_{g,f}\left(x_0,x\right)=\mathrm{e}^{-\int_{x_0}^x \frac{1}{u_g \tau_g\left(x^{\prime}\right)} \mathrm{d} x^{\prime}}.
\end{equation}
Conversely, during transport, the probability density function (PDF) of experiencing a collision at point $x^\prime$ and keeping the local equilibrium distribution to $x$ is:
\begin{equation}
P_{g,c}\left(x^\prime,x\right)=\frac{1}{u_g \tau_g\left(x^{\prime}\right)} \mathrm{e}^{-\int_{x^{\prime}}^x \frac{1}{u_g \tau_g\left(x^{\prime \prime}\right)} \mathrm{d} x^{\prime \prime}}.
\end{equation}
Considering the composition of the distribution function at a certain point $x$, the final distribution function at a certain point $x$ can be regarded as the mathematical expectation of the distribution functions from the starting point $x_0$ to that point $x$, weighted by their respective probabilities.

In other words, the distribution function at that point $x$ is a convex combination of the equilibrium distribution functions from itself and other positions and the boundary distribution function.
Suppose the boundary distribution function is at equilibrium. In that case, the distribution function at that point is a convex combination of the equilibrium distribution functions from all the points in the computational region:
\begin{equation}\label{sum-g}
    e_{g,i, k}^{n+1} = \sum_{j=0}^{N}\omega_{g,j,k}e^{eq,n}_{g,j,k},
\end{equation}
where N is the total number of cells of the grid.
Considering the case where $\tau_g$ is constant, and based on previous results that yield $P_{g,c}=\frac{dP_{g,f}}{dx}$, we can further rewrite the above equation as:
\begin{equation}\label{sum-g-CDF}
    e_{g,i, k}^{n+1}\left(x,u_g\right) = \sum \mathbf F\left({P_{g,f}\left(x_j,x\right)}\right)e^{eq,n}_{g,j,k}\left(x_j,u_g\right),
\end{equation}
where the summation takes over all particles with the specified velocity capable of traveling from point $x_j$ to point $x$ and $\mathbf{F}$ is the function related to the CDF.
This indicates that the distribution function at point x is obtained by probabilistically accumulating the corresponding equilibrium distribution functions from other points using the CDF.

\subsection{Particle evolution in IUGKP}
The focus of this paper is the multi-frequency method under frequency-dependent conditions. 
Therefore, in this section, we only present the algorithm formulation for the case where $\tau_g$ is constant.
For the case where $\tau_g$ varies with space, please refer to our previous work \cite{liu2025implicit}.

From the emission perspective, particles emitted from a given point in local equilibrium have free transport distances that follow the cumulative PDF $P_{g,f}$.
From the inverse sampling method, its free transport length is:
\begin{equation}
    \lambda_g = -u_g\tau_g ln\left(\eta\right).
\end{equation}
After traveling this distance, the particle undergoes a collision at the local point and then conforms to the local equilibrium distribution.
Then, Eq.~\eqref {sum-g-CDF} can be reformulated into a particle-based representation as follows:
\begin{equation}\label{sum-g-particle}
    Eg,_{i} = \sum w_g^p.
\end{equation}

Therefore, for a constant value of $\tau$, the steps for one iteration of the steady-state algorithm are:

(1) for each particle in the cell, resampling its distribution, which satisfies the local equilibrium state.

(2) sampling all the particles' free streaming path and moving them to the target cell according to this length.

(3) Update the cell average macro variables by counting the masses, momentum, and energy of the particles in the cell.

\begin{remark}
The gray model for phonon transport can be regarded as a multi-group phonon transport equation with only one frequency, whose compatibility condition is shown as Eq~.\eqref{gray-capability} and whose equilibrium state corresponds to an energy equal to the total energy shown as Eq~.\eqref{gray-conservation}
\begin{equation}\label{gray-capability}
    \iint_{4\pi}\frac{e^{eq}_g-e_g}{\tau_g} \mathrm{~d} \Omega= 0,
\end{equation}
\begin{equation}\label{gray-conservation}
    E_g=C_g\left(T-T_{ref}\right).
\end{equation}
However, the compatibility conditions for the multi-group transport equations differ from those of the gray model; that is, as mentioned above, the angular integration of the collision operator for each group is not zero. In other words, the energy associated with each group is not equal to the energy corresponding to its equilibrium state, shown as Eq~.\eqref{multi-capability-unequal}
\begin{equation}\label{multi-capability-unequal}
    \iint_{4\pi}\frac{e^{eq}_g-e_g}{\tau_g} \mathrm{~d} \Omega \neq 0.
\end{equation}
\end{remark}

\begin{remark}
The particle sampling at the new iteration step should follow the equilibrium distribution of the current group, as shown in Eq.~\eqref {multi-group-sample}, rather than Eq.~\eqref {gray-group-sample}. Eq.~\eqref {gray-group-sample} corresponds to the sampling rule of the gray model.
\begin{equation}\label{multi-group-sample}
    \sum w_g^{s} = \iint_{4\pi} e^{eq}_g \mathrm{~d} \Omega
\end{equation}\label{gray-sample}
\begin{equation}\label{gray-group-sample}
    \sum w_g^{s} = E_g
\end{equation}
\end{remark}

\subsection{Adaptive sampling of particles in frequency space}
In multi-group phonon transport, if Monte Carlo sampling is performed only in the angular space, it is equivalent to converting the two-dimensional discrete velocity space into a two-dimensional Monte Carlo sampling, and the frequency space is still handled by applying a discrete velocity method to solve the BGK equation at each discrete frequency interval.
To further accelerate computational speed, this section introduces adaptive sampling of particles in frequency space.

Since the energy-form phonon transport BGK equation is adopted in this paper, the functions used to determine the particle interval assignment should also be in the energy form.
The particle energy corresponding to the equilibrium state contained in each frequency interval can be expressed by Eq~.\eqref{energy-each-group}:
\begin{equation}\label{energy-each-group}
    E_{g}^{eq}=\iint_{4\pi} e^{eq}_g\mathrm{~d}\Omega\Delta\omega_g= C_g\left(T-T_{ref}\right)\Delta \omega_g.
\end{equation}

Therefore, the energy fraction for each phonon group is given as shown in Eq~.\eqref{energy-fraction}:
\begin{equation}\label{energy-fraction}
    F_g=\frac{C_g\left(T-T_{ref}\right)\Delta \omega_g}{\sum_p \sum_gC_g\left(T-T_{ref}\right)\Delta \omega_g}=\frac{C_g\Delta \omega_g}{\sum_p \sum_gC_g\Delta \omega_g}.
\end{equation}

Thus, the cumulative distribution function of energy fraction $H_g$ can be written as Eq~.\eqref{Hg}:
\begin{equation}\label{Hg}
H_g=\sum_0^{g}F_g.
\end{equation}

Then, given a random number $\eta$ which follows a uniform distribution between 0 and 1, the particle belonging to the $g^{th}$ frequency interval, which satisfies:
\begin{equation}\label{frequency-determine}
    H_{g}>\eta >H_{g-1},
\end{equation}
where $H_{-1}=0$.

For IUGKP, the number of particles at each frequency remains fixed throughout the evolution process, so the particle number for each frequency only needs to be determined initially, and the specific sampling method for each grid cell is as follows:

(1) Give the reference sampling number $N_{ref}$.

(2) Based on Eq~.\eqref{energy-fraction} and Eq~.\eqref{Hg}, calculate the cumulative distribution function of energy fraction for each group.

(3) Sample $N_{ref}$ particles, and for each particle, determine its corresponding frequency interval according to Eq~.\eqref{frequency-determine}.

(4) If the number of sampled particles for the $g^{th}$ frequency is $N_g$, and this number is less than the specified minimum value $N_{min}$, then an additional $(N_{min} - N_g)$ particles should be sampled for that frequency.

After performing the above Monte Carlo sampling in frequency space, the total number of particles can be reduced from the order of $N_B \times N_{ref}$ to the order of $N_{ref}$, significantly saving overall memory consumption and improving computational speed.

The core idea behind this method is that phonon groups with higher energy fractions contribute more to the final energy, and therefore, they require more discrete velocity-space resolution (or more statistical particles) to capture their evolution accurately.
For phonon groups with small contributions, such as those accounting for only $0.01\%$, regardless of their Knudsen number, their impact on the final energy is negligible. 
Therefore, they require only a very small amount of velocity space or statistical particles to maintain computational stability.

The concept of adaptive frequency-space sampling can also be applied to the UGKWP method, although its specific implementation differs. For UGKWP, the reference sampling number of particles at each frequency point can be allocated according to the following formula:
\begin{equation}
    N_{g,ref}=F_g\times N_{ref},
\end{equation}
where $N_{g,ref}$ is the reference sampling number of particles for $g^{th}$ group.
This also adheres to the principle that frequencies with a higher energy proportion are allocated more sampling particles.

\subsection{Macro prediction}

In the current particle evolution method, the evolution scale is determined by the physical mean free path. 
Therefore, at high Knudsen numbers the method converges quickly because particles travel relatively longer distances between collisions~\cite{ADAMS02fastiterative}. 
However, at small Knudsen numbers, the mean free path becomes shorter, which in turn significantly slows down the convergence speed of the method.
Meanwhile, in the case of small Knudsen numbers, macroscopic equations boast extremely high efficiency. 
Therefore, constructing macroscopic equations to accelerate convergence is a feasible approach.
This section will discuss how to construct macroscopic equations to accelerate convergence when the Knudsen number is small.

For the sake of clarity in the subsequent discussion, we first review the inexact Newton iteration method~\cite{Chuang17gray,zhang2019implicit}.
To solve $F\left(x\right)=0$, inexact Newton method uses a approximate operator $J\left(\delta x\right)$ to solve the equation~\cite{Chuang17gray,zhang2019implicit}
\begin{equation}\label{inexact-newton}
    J\left(\delta x^k\right)=-F\left(x^k\right),
\end{equation}
after $k^{th}$ step iteration,
\begin{equation}
    x^{k+1}=x^{k}+\delta x^{k}.
\end{equation}
As iteration converges, $x^{k+1}=x^{k}$ and $\delta x=0$, then we have:
\begin{equation}\label{inexact-newton}
   F\left(x^k\right)= -J\left(\delta x^k\right)=0.
\end{equation}

For multi-frequency phonon transport, the corresponding macroscopic equation without the external heat source is the conservation of total heat flux:
\begin{equation}
    \nabla \cdot \boldsymbol{q}=0,
\end{equation}
where $\boldsymbol{q}$ can be obtained by the particle flux:
\begin{equation}\label{particle-res}
    \nabla \cdot \boldsymbol{q} = -\frac{\sum w_p^{in} - \sum w_p^{out}}{V},
\end{equation}
where $w_p^{in}$ is the energy of the particle moving into the target cell, $w_p^{out}$ is the energy of the particle leaving the target cell. 

The Fourier heat conduction equation can be viewed as an approximate operator for the conservation of heat flux, similar to the $J$ operator, so we can construct an inexact Newton iteration as:
\begin{equation}\label{inexact-newton-heat}
    -\nabla \cdot\left(\kappa_{eff} \nabla \delta T\right) = \nabla \cdot \boldsymbol{q},
\end{equation}
where $\kappa_{eff}$ is the effective thermal conductivity, and its expression is given by Eq~.\eqref{keff}
\begin{equation}\label{keff}
    \kappa_{eff}=\sum_p\int_{\omega_{min}}^{\omega_{max}} \frac{C_g\tau_g|\boldsymbol{V_g}|^2}{3} \mathrm{~d}\omega.
\end{equation}

In practical calculations, an amplification factor $\gamma$ is applied to the effective thermal conductivity. 
The purpose of this factor is to reduce the effect of statistical noise in the net particle flux on the right-hand side, thereby stabilizing the computation. 
Consequently, Eq~.\eqref{inexact-newton-heat} is modified into Eq~.\eqref{inexact-newton-heat-new}
\begin{equation}\label{inexact-newton-heat-new}
    -\nabla \cdot\left(\gamma\kappa_{eff} \nabla \delta T\right) = \nabla \cdot \boldsymbol{q}.
\end{equation}
Then the predicted temperature for sampling the particles in the next iteration step is:
\begin{equation}
    \tilde{T}=T+\delta T.
\end{equation}
\begin{remark}
Moving the effective thermal conductivity to the right-hand side, we have:
    \begin{equation}\label{change-heat-equation}
        -\nabla \cdot\left( \nabla \delta T\right) = \frac{\nabla \cdot \boldsymbol{q}}{\kappa_{eff}}.
    \end{equation}
When the Knudsen number is small, the effective thermal conductivity is also small. 
The right-hand side of the Eq~.\eqref {change-heat-equation} creates a source term with a high degree of stiffness.
Therefore, we introduce the amplification factor to reduce the statistical noise and mitigate the large stiffness caused by small effective thermal conductivity, making the computation more stable.
\end{remark}

\begin{remark}
For the sake of clarity, without loss of generality, considering the 1D uniform grid, then Eq~.\eqref{change-heat-equation} can be further written as:
    \begin{equation}\label{change-heat-equation-1d}
        \delta T^{n+1}_{i+1} - 2\delta T^{n+1}_{i}+\delta T^{n+1}_{i-1} = \left(\sum w_p^{in} - \sum w_p^{out}\right) \frac{\Delta x}{\sum_p \int_{\omega_{\min }}^{\omega_{\max }}\frac{C_g \tau_g\left|\boldsymbol{V}_g\right|^2}{3} \mathrm{~d} \omega}.
    \end{equation}
It can be simplified as:
    \begin{equation}\label{change-heat-equation-1d-sp}
        \delta T^{n+1}_{i+1} - 2\delta T^{n+1}_{i}+\delta T^{n+1}_{i-1} = \frac{\left(\sum w_p^{in} - \sum w_p^{out}\right)}{N_{cell}} \frac{L_{ref}}{\sum_p \int_{\omega_{\min }}^{\omega_{\max }}\frac{C_g \tau_g\left|\boldsymbol{V}_g\right|^2}{3} \mathrm{~d} \omega},
    \end{equation}
where,$\Delta x=\frac{L_{ref}}{N_{cell}}$, $L_{ref}$ is the characteristic length.
when the characteristic length goes to zero, it means the $\text{Kn}_g$ goes to infinity, and $\text{Kn}_g$ is the Knudsen number for the $g^{th}$ group, meaning the transport is nearly ballistic.
As a result, the right-hand side of the Eq~.\eqref {change-heat-equation-1d-sp} goes to 0.
Thus, we are solving:
    \begin{equation}\label{change-heat-equation-big-kn}
        \delta T^{n+1}_{i+1} - 2\delta T^{n+1}_{i}+\delta T^{n+1}_{i-1} = 0,
    \end{equation} with the boundary condition: $\delta T_{ghost}=0$.
    Solving such an equation, we obtain $\delta T=0$.
    This indicates that the macroscopic prediction equation will play a little role in the ballistic transport limit, which is the intended outcome of this manuscript, because in the ballistic transport regime, particle evolution is highly efficient. The same applies to high-dimensional cases.
\end{remark}

Algorithm~.\ref{iugkp-algorithm} presents a summary of the prediction-correction IUGKP method for frequency-dependent phonon transport:

\begin{algorithm}\label{iugkp-algorithm}
	\caption{IUGKP method for frequency-dependent phonon transport}
	\label{wp-algorithm}
	Initialization: set the initial temperature and energy field for each group\;
	  \textbf{while} (not converged) \textbf{do}\;
        Solve the macro prediction equation to obtain $\tilde{T}$\;
        Resample particles' energy and velocity\;
        Move all the particles to their destination\;
        Calculate the total energy for each phonon group\;
        Update the total heat flux by Eq~.\eqref{particle-res}\;
        Update temperature by Eq~.\eqref{temperature-update}\;
        \textbf{end while}\;
        Output
\end{algorithm}	

\section{Numerical Tests}
In this section, silicon is used as an example to calculate the multiscale heat conduction in actual materials and to compare the results with existing reference solutions.
Taking the $\left[1,0,0\right]$ direction of the silicon as the representative direction, it is assumed that the dispersion relations in all other directions are identical to that of this direction.
Silicon has three acoustic branches: two longitudinal acoustic (LA) branches and one transverse acoustic (TA) branch. Since the optical branches have a negligible effect on thermal conductivity, they are neglected.
This section uses a series of benchmark heat transfer cases to validate the effectiveness of the proposed multiscale method for solving the phonon BTE.
In this section, a quadratic polynomial is used to approximate silicon's dispersion relation \cite{zhang2019implicit}:
\begin{equation}
    \omega=10^3c_1 k+10^{-7}c_2 k^2,
\end{equation}
where $c_1$ and $c_2$ are coefficients and their value are listed in Table~.\ref{c1-and-c2}
\begin{table}[htb]
	\small
	\begin{center}
		\def\temptablewidth{1.0\textwidth}
		{\rule{\temptablewidth}{1pt}}
		\begin{tabular*}{\temptablewidth}{@{\extracolsep{\fill}}c|c|c}
			phonon branch & $c_1$ & $c_2$ \\
			\hline
			LA 	&9.01 & -2.0  \\ 	
                TA 	&5.23 & -2.26   \\
		\end{tabular*}
		{\rule{\temptablewidth}{0.1pt}}
	\end{center}
	\vspace{-4mm} \caption{\label{c1-and-c2} Coefficients of the dispersion relationship quadratic polynomial.}
\end{table}

The wave vector $k$ belongs to the interval $[0,k_{max}]$, and $k_{max}$ is the maximum wave vector in the first Brillouin zone:
\begin{equation}
    k_{\max }=10^{10}2 \pi / a.
\end{equation}
where $a$ is the lattice constant.  For silicon, $a=5.43 \AA$.

Then the phonon group velocity can be obtained by:
\begin{equation}
    \boldsymbol{V_g}=|\boldsymbol{V_g}| \boldsymbol{s}=\left(10^3c_1+2 \times 10^{-7} c_2 k\right) \boldsymbol{s} .
\end{equation}

As for relaxation time, this paper adopted Terris' rule \cite{terris2009modeling}:
\begin{equation}
\tau^{-1}=\tau_{\text {impurity }}^{-1}+\tau_{\mathrm{U}}^{-1}+\tau_{\mathrm{N}}^{-1}=\tau_{\text {impurity }}^{-1}+\tau_{\mathrm{NU}}^{-1}
\end{equation}
The formulas and coefficients for the relaxation time are given in Table~.\ref{tab:relaxation_time}
\begin{table}[h]
    \centering
    \begin{tabular}{cl}
        \hline
        $\tau_{\text{impurity}}^{-1}$ &  $A_t \omega^4, \, A_t = 1.498 \times 10^{-45} \, \text{s}^3;$\\
        \hline
        LA & $\tau_{\text{NU}}^{-1} = B_L \omega^2 T^3, \, B_L= 1.180 \times 10^{-24} \, \text{K}^{-3};$ \\
        TA & $\tau_{\text{NU}}^{-1} = B_T \omega^4, \, 0 \leq k < k_{\text{max}}/2;$ \\
        & $\tau_{\text{NU}}^{-1} = B_U \omega^2 / \sinh(\hbar \omega / k_B T), \, k_{\text{max}}/2 \leq k < k_{\text{max}};$ \\
        & $B_T = 8.708 \times 10^{-13} \, \text{K}^{-3}, \, B_U = 2.890 \times 10^{-18} \, \text{s}.$ \\
        \hline
    \end{tabular}
    \caption{Relaxation Time Expressions for LA and TA Phonons}
    \label{tab:relaxation_time}
\end{table}

\begin{figure}[htp]
	\centering	
	\includegraphics[height=0.40\textwidth]{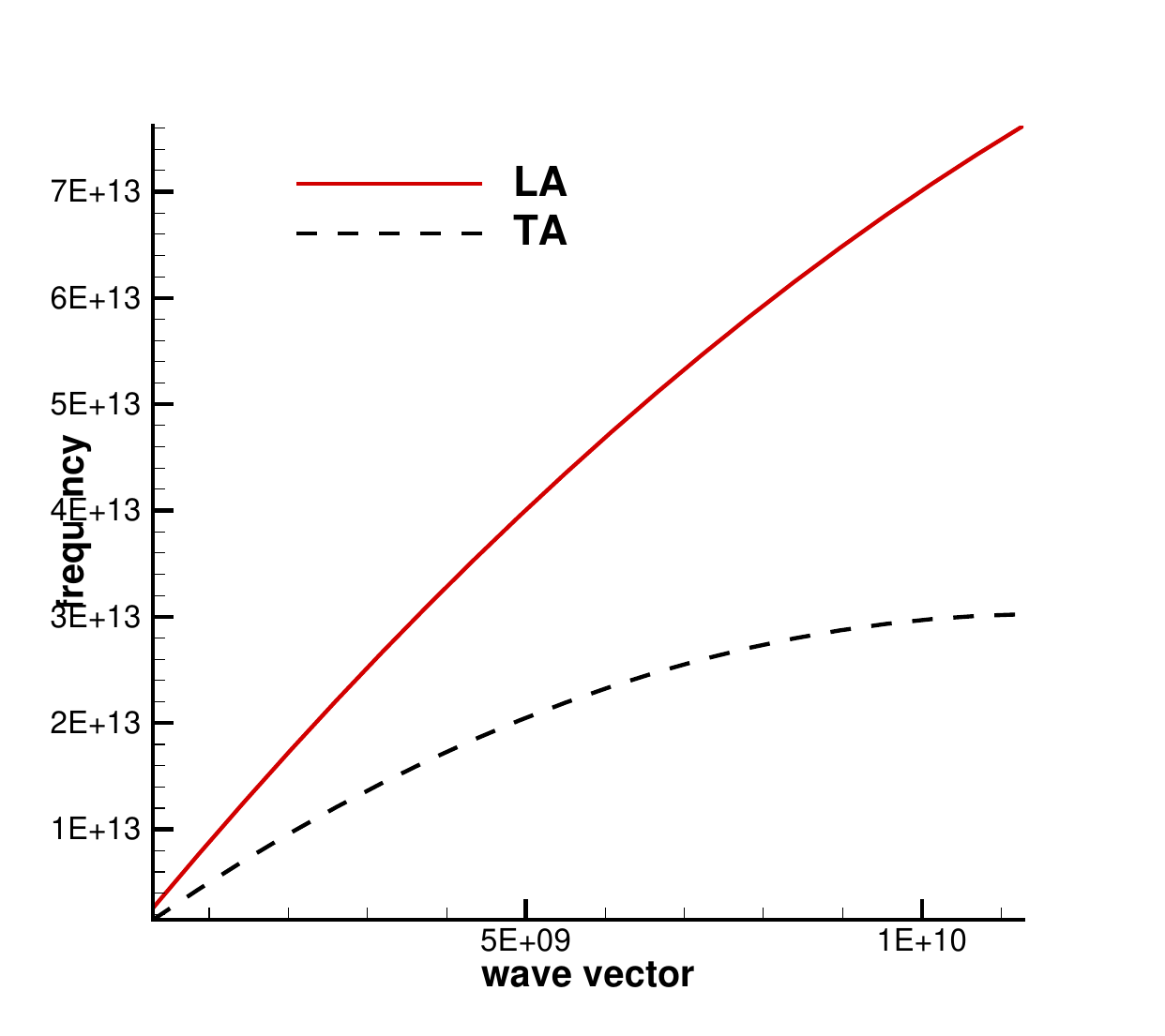}
	\includegraphics[height=0.40\textwidth]{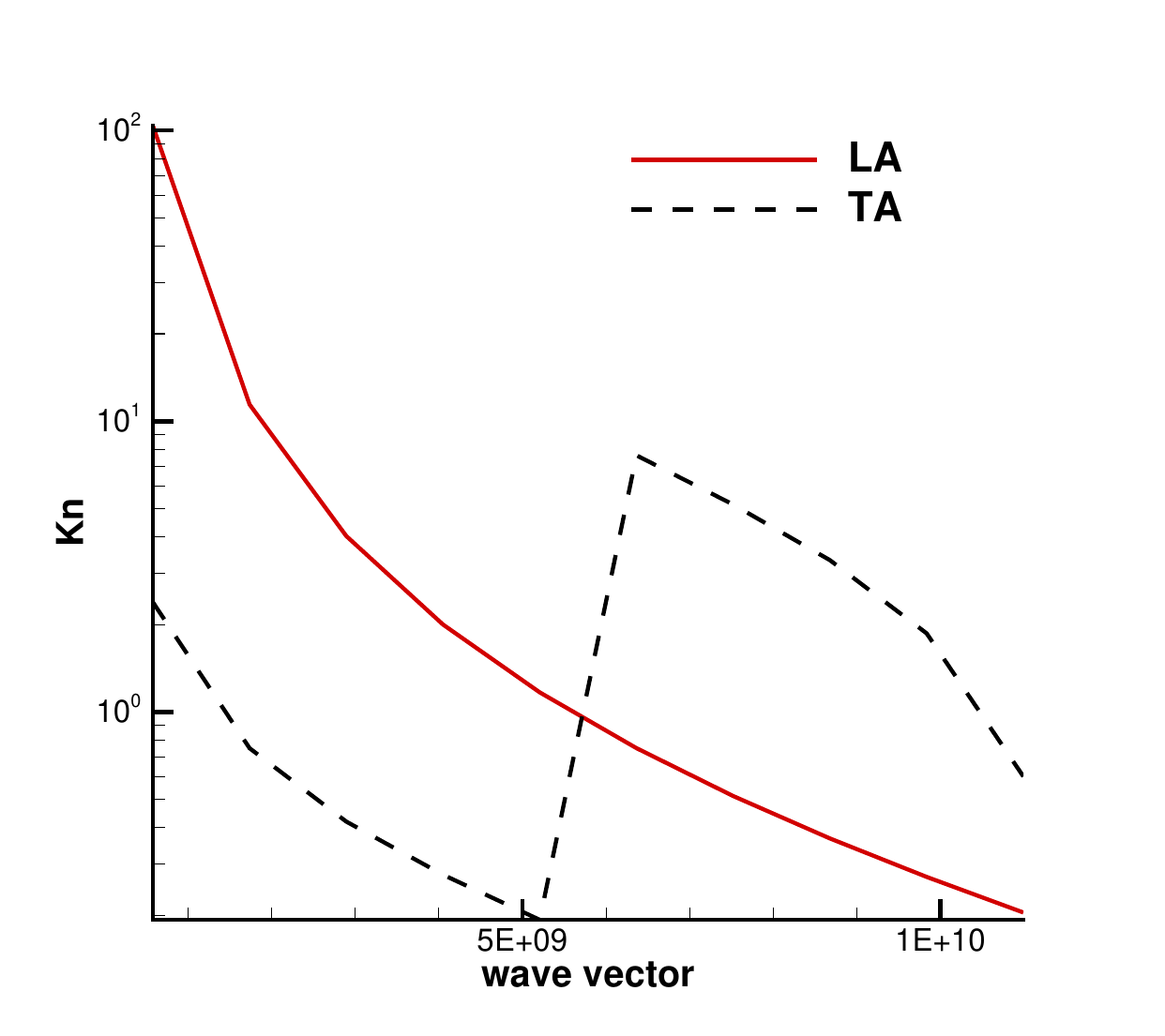}
	\caption{\label{dispersion-relationship}
		Dispersion relationship.}
\end{figure}

The dispersion line and the Kn number at different frequencies are shown in Fig~.\ref{dispersion-relationship}

The time step for UGKWP is determined by $\Delta t=\mathrm{CFL} \times \frac{\Delta x}{max\boldsymbol{|V_g}|}$ with $\mathrm{CFL}=0.8$.
The results obtained by the present method will be
compared with the data predicted by the DUGKS \cite{zhang2021fast}.
The van Leer limiter is used for UGKWP to ensure numerical stability.
Moreover, all the numerical tests use $N_B=20$ points for discretizing the frequency space of each phonon branch.

A series of numerical tests have been performed to verify the current schemes' accuracy and efficiency. 
All the numerical tests, including large-scale three-dimensional multiscale heat conduction, are run on a personal laptop using a single core, which is nearly impossible for the DOM method, and the CPU is \text { 12th Gen Intel(R) Core(TM) i7-1260P }.

\subsection{1D cross-plane heat conduction}
In this section, one-dimensional heat conduction in a dielectric film is simulated for three cases: $L = 10$ nm, $L = 100$ nm, $L = 1000$ nm, and $L = 100000$ nm as illustrated in Fig~.\ref{1d-heat}.

\begin{figure}[htb]	\label{1d-heat}
	\centering	
	\includegraphics[height=0.45\textwidth]{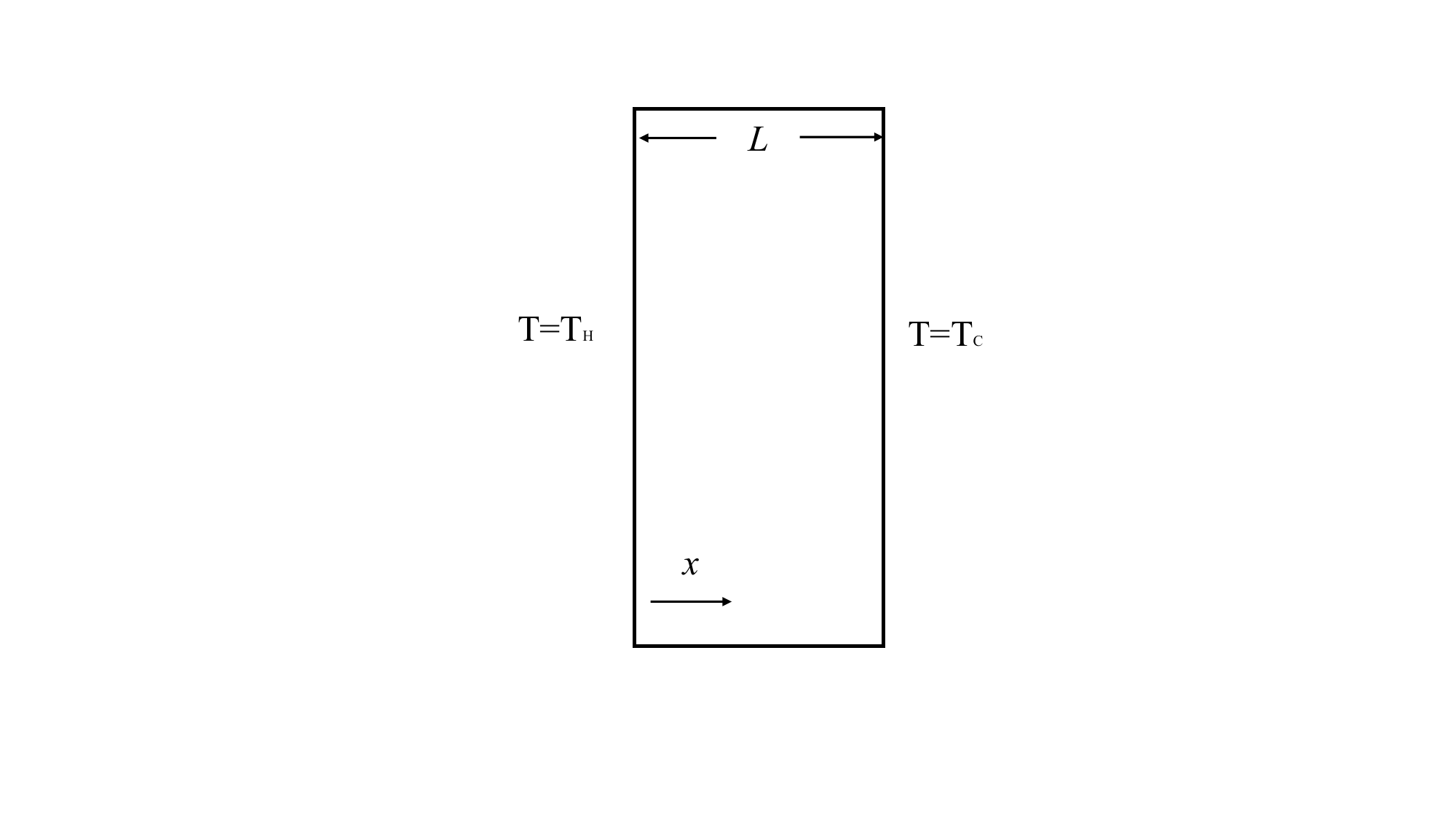}
	\caption{\label{1d-heat}
		A schematic diagram of one-dimensional heat conduction in a dielectric film.}
\end{figure}
At the left boundary $(x = 0)$, an isothermal high-temperature boundary is imposed with a temperature of $T_H$, while at the right boundary $(x = L)$, an isothermal low-temperature boundary is imposed with a temperature of $T_L$.
The reference solutions can be obtained in \cite{zhang2021fast}.

To minimize statistical noise, the reference number for sampling particles in each cell is set to 100,000.
The one-dimensional computational domain is discretized into 40 uniform cells, and the CFL number is set to 0.8.

Fig.\ref{1d-heat-result} illustrates a comparison among the computed results of the IUGKP method, the UGKWP method, and the reference solution, where $T^*=\frac{T-T_C}{T_H-T_C}$, $x^*=\frac{x-x_L}{x_R-x_L}$.
The results indicate that the IUGKP method and the UGKWP method agree well with the reference solution. 
Moreover, we use a larger number of particles in the one-dimensional case to reduce statistical noise rather than relying on statistical averaging. 
Thus, we compare only the number of iteration steps required for convergence between IUGKP and UGKWP in this example, rather than comparing computational time. 
This approach reflects the accelerated convergence effect of the IUGKP method on steady-state phonon transport. 
The differences in computational time will be compared in the subsequent two-dimensional examples.

As shown in Table~.\ref{1d-compare-iugkp-vs-ugkwp}, IUGKP achieves a 1 to 2 order of magnitude improvement in convergence speed compared to UGKWP, from the diffusion limit to the ballistic transport limit.

\begin{table}[htb]
	\small
	\begin{center}
		\def\temptablewidth{1.0\textwidth}
		{\rule{\temptablewidth}{1pt}}
		\begin{tabular*}{\temptablewidth}{@{\extracolsep{\fill}}c|c|c|c|c}
			Scheme & L = 10nm & L = 100nm  & L = 1$\mu$m  & L = 100$\mu$m  \\
			\hline
			UGKWP 	&1600 & 1800  & 4000  & 12,0000  \\ 	
                IUGKP 	&4    & 10    & 200   & 320 \\
                Ratio 	&400  & 180   & 20    & 38 \\
		\end{tabular*}
		{\rule{\temptablewidth}{0.1pt}}
	\vspace{-4mm} \caption{\label{1d-compare-iugkp-vs-ugkwp} Iteration steps of UGKWP and IUGKP for 1D cross-plane heat conduction.}
	\end{center}
\end{table}

\begin{figure}[htb]	\label{1d-heat-result}
	\centering	
        \includegraphics[height=0.40\textwidth]{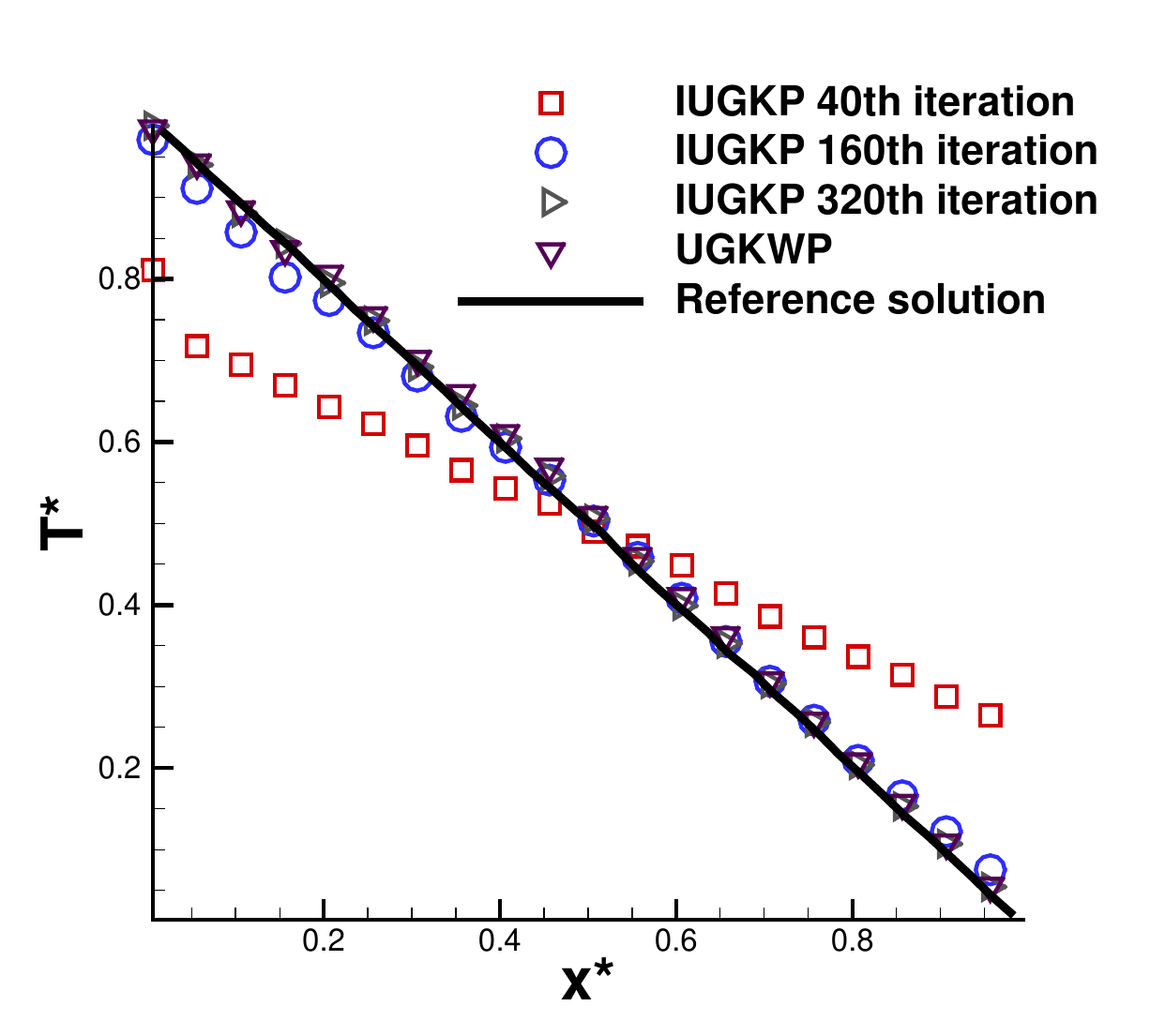}
	\includegraphics[height=0.40\textwidth]{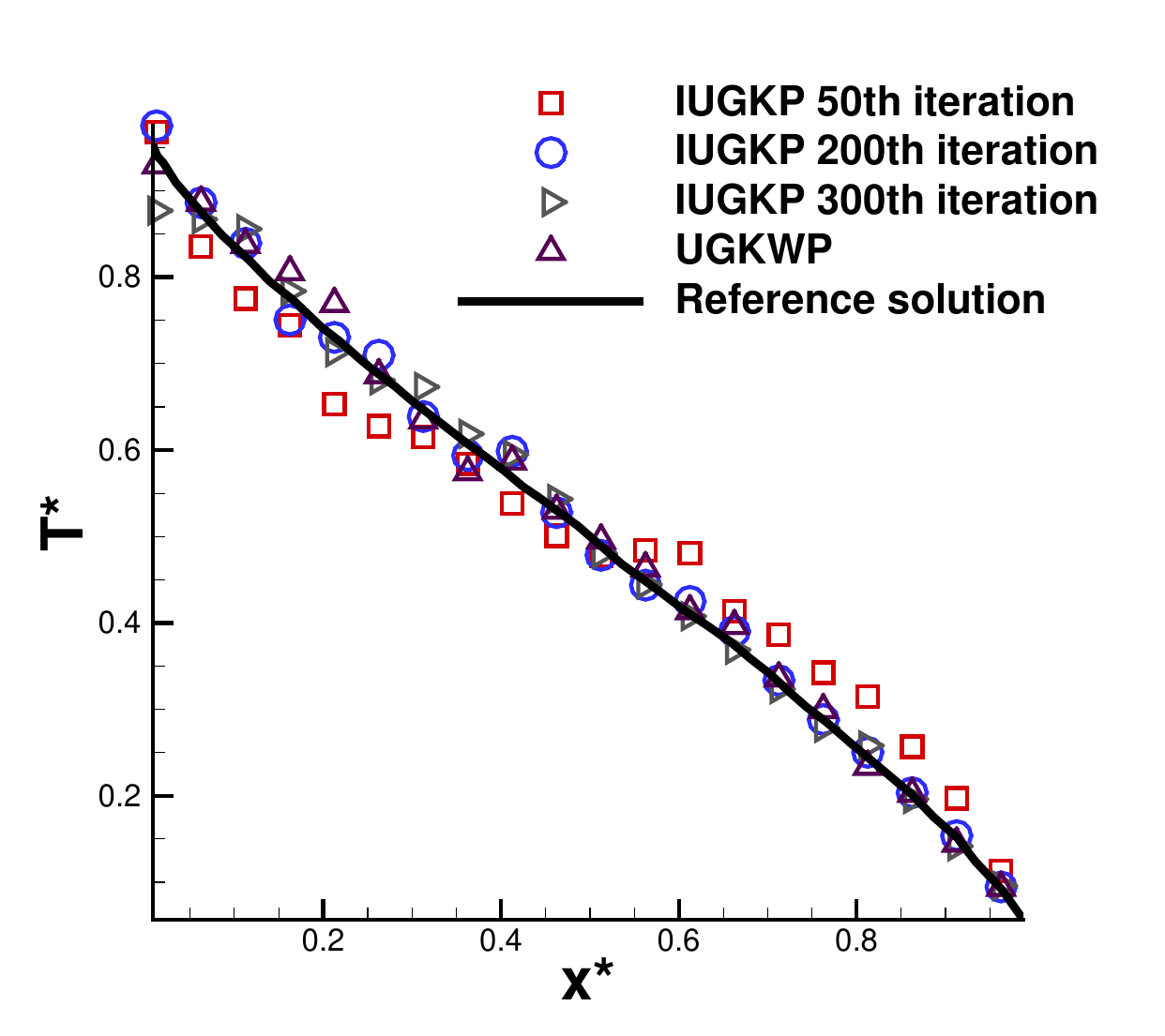}
        \includegraphics[height=0.40\textwidth]{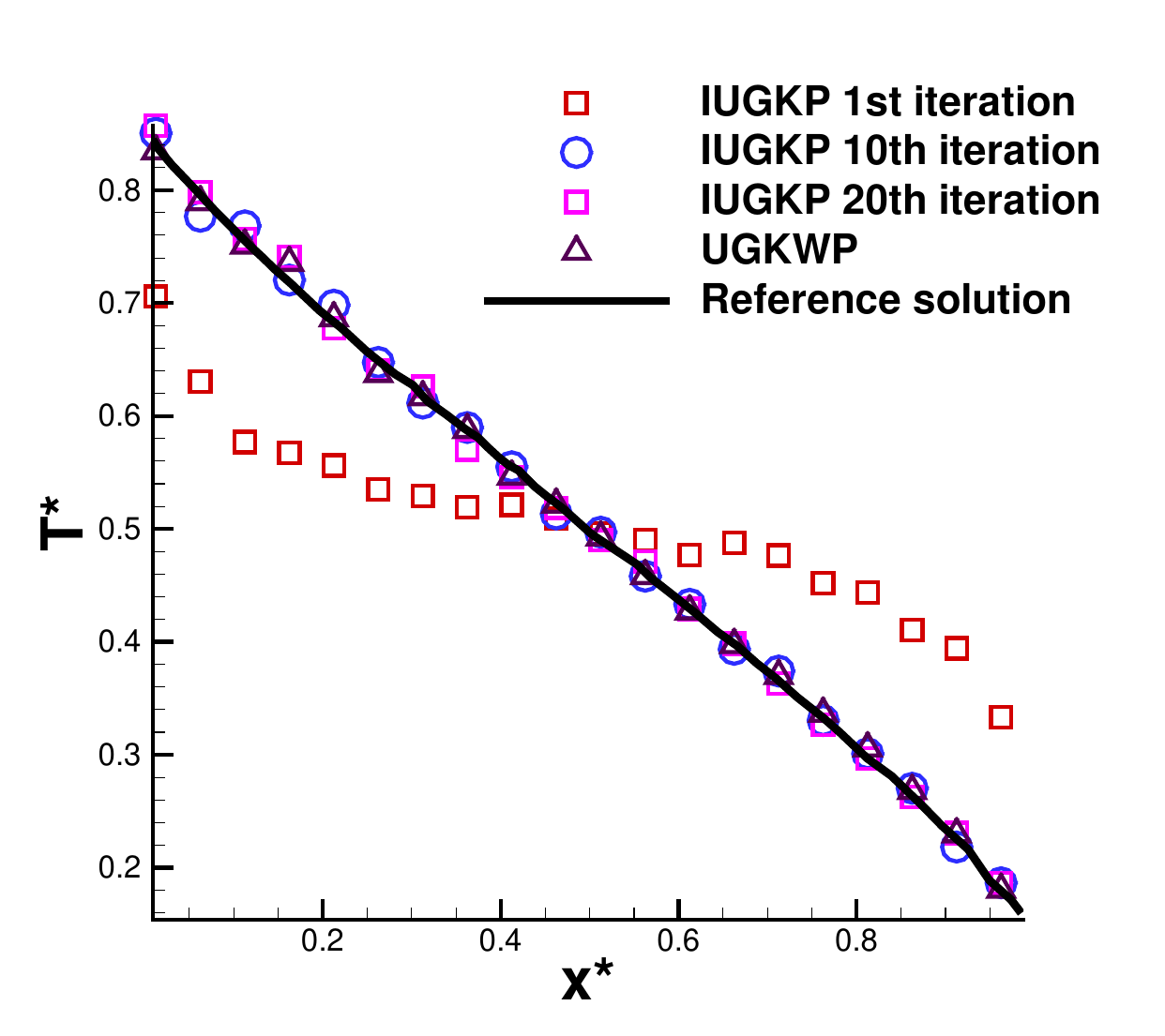}
        \includegraphics[height=0.40\textwidth]{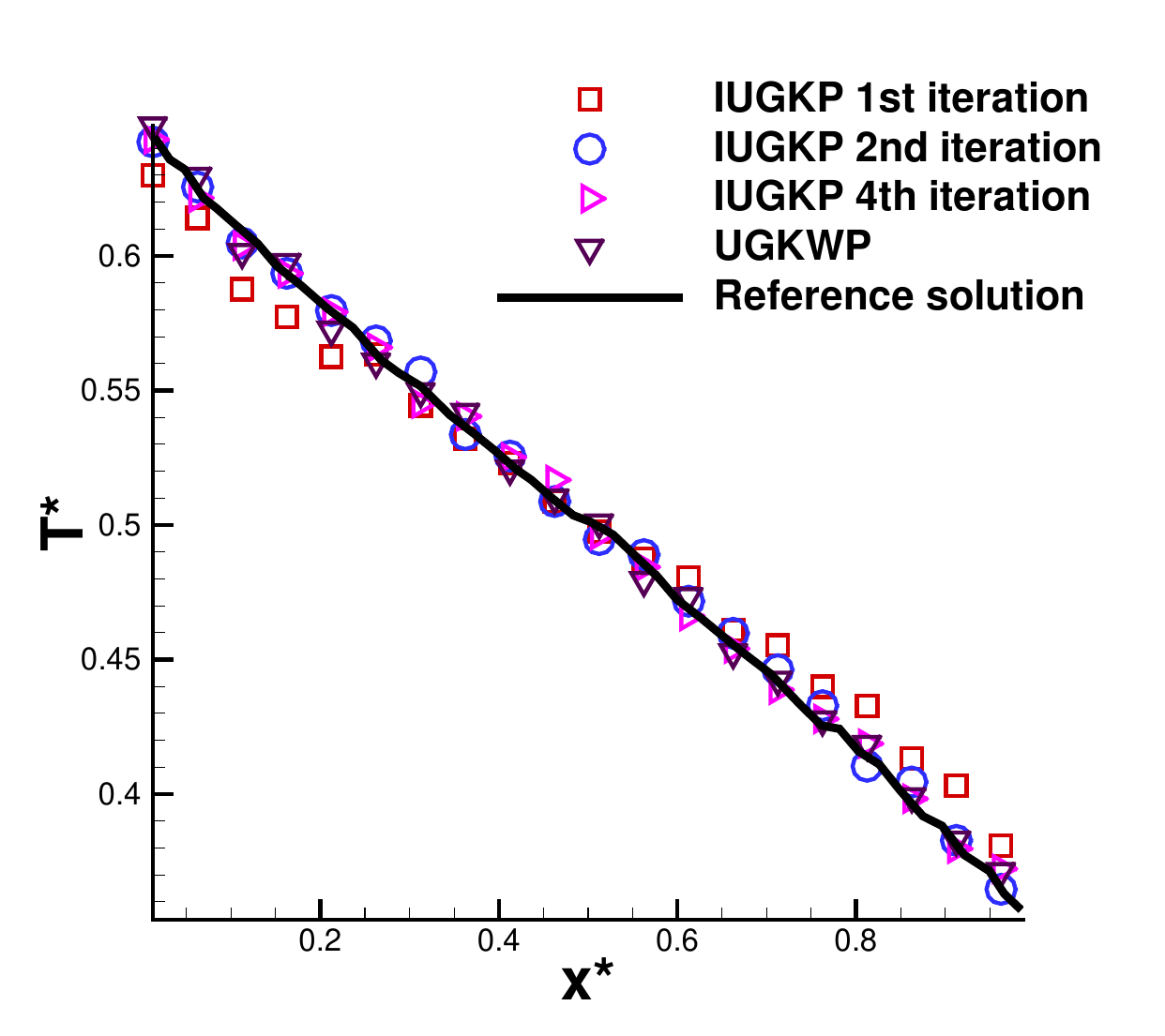}
	\caption{\label{1d-heat-result}
		Comparison of 1D heat conduction results across a ﬁlm at $L=100\mu m, 1 \mu m, 100 nm$, $10 nm$ (left to right, and up to down). }
\end{figure}

\subsection{1D transient cross-plane heat conduction}
To verify the UGKWP method for solving unsteady cases, this section tests the 1D transient cross-plane heat conduction for $L$ = 10, 100, and 1000 nm.
The setup for the numerical test is identical to that in the previous section, and the reference solution is obtained using the UGKS method, in which 200 discrete velocity points are employed.
The computational results are presented in Fig~.\ref{1d-transient-heat-result}, which agree well with the reference solution, demonstrating the accuracy of the current UGKWP method in simulating unsteady problems.

\begin{figure}[htb]	\label{1d-transient-heat-result}
	\centering	
        \includegraphics[height=0.35\textwidth]{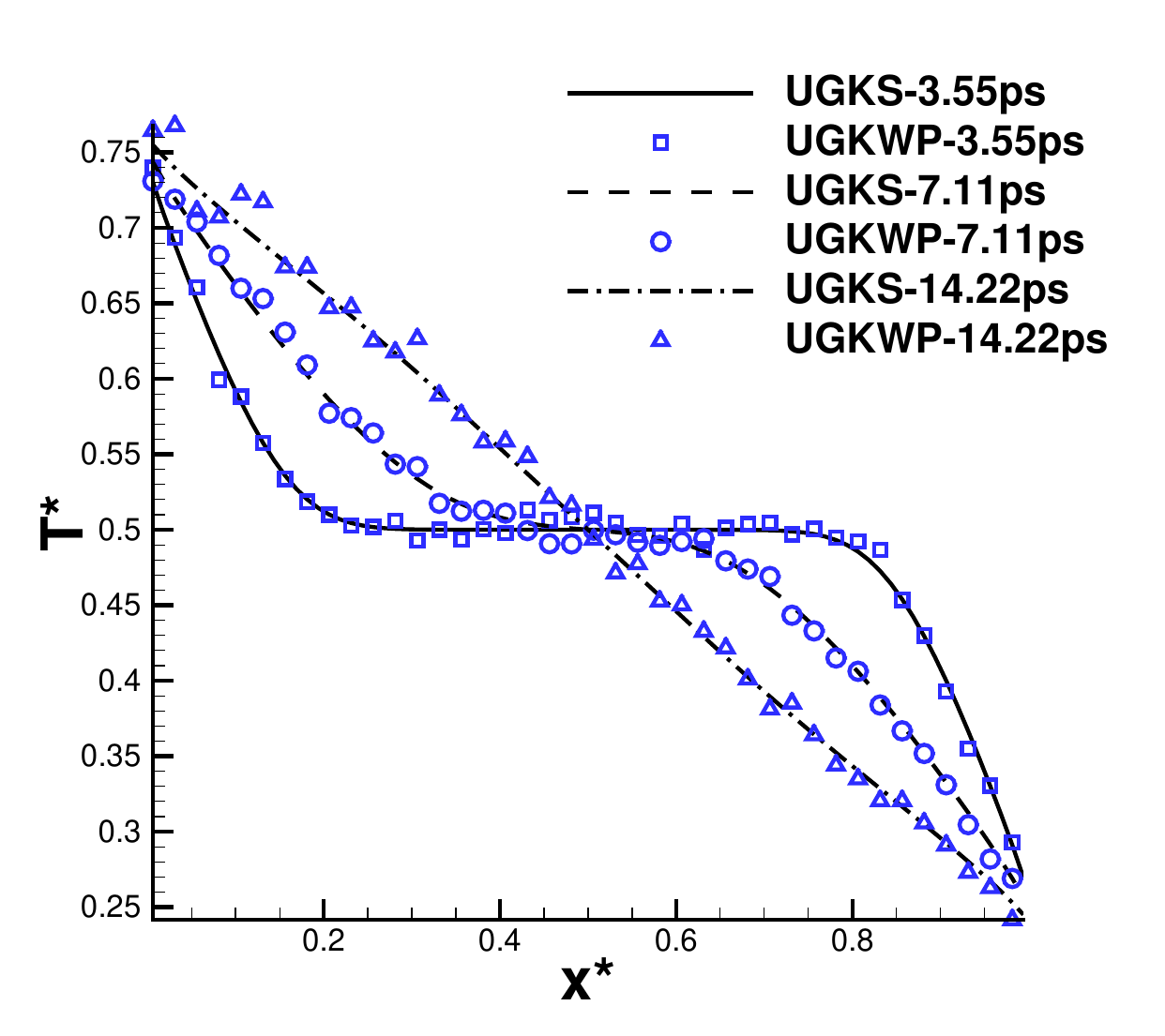}
	\includegraphics[height=0.35\textwidth]{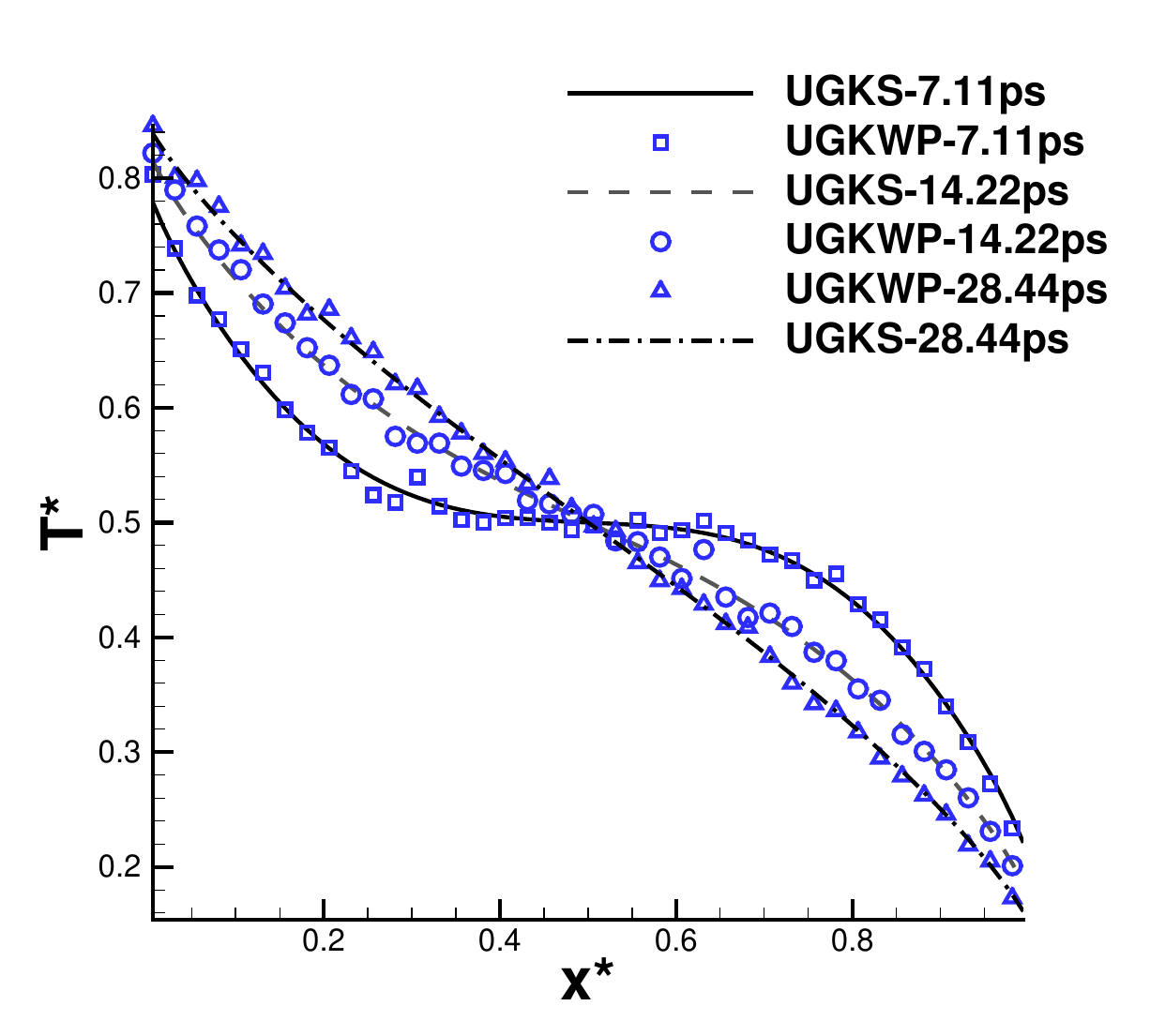}
        \includegraphics[height=0.35\textwidth]{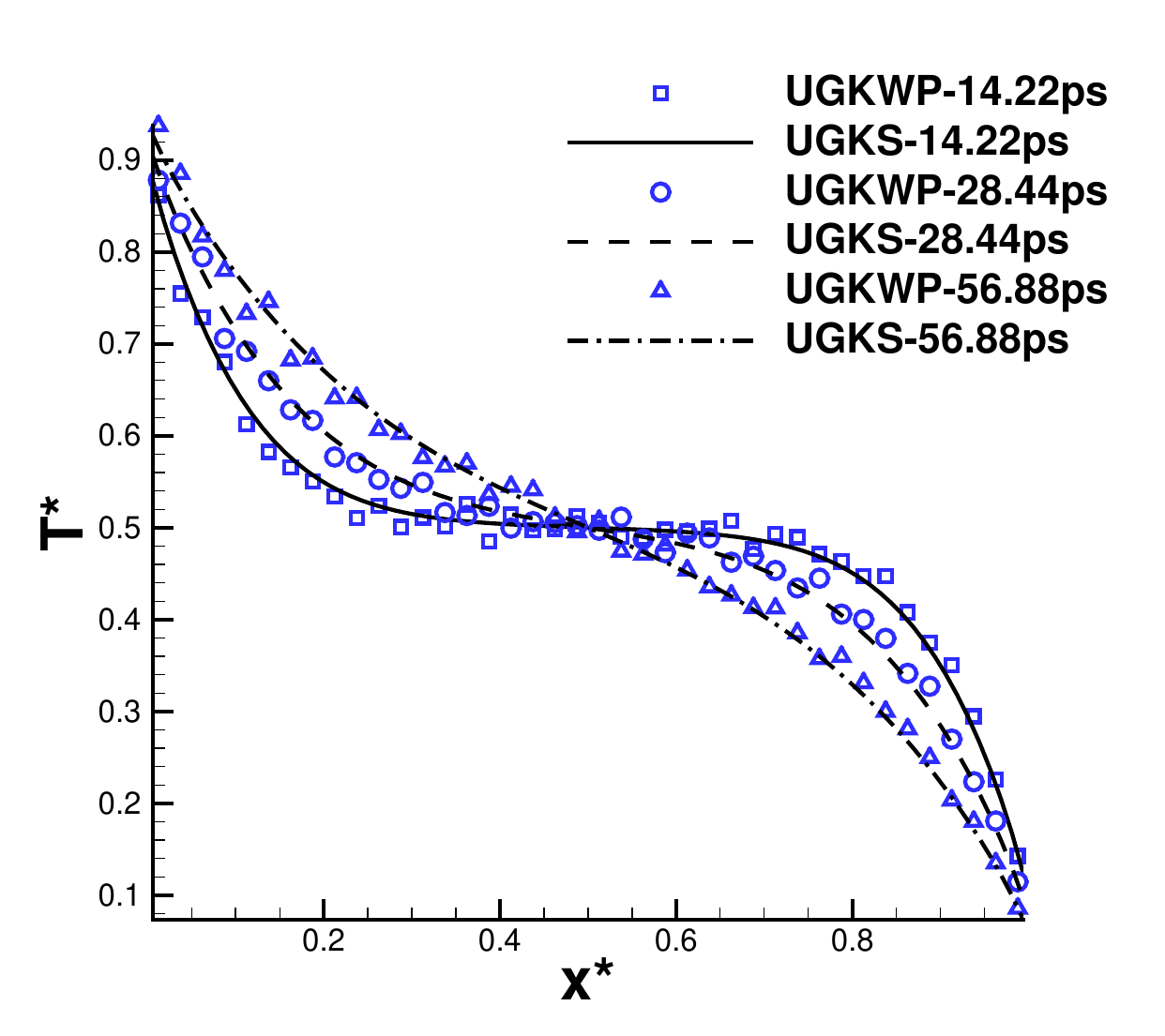}
	\caption{\label{1d-transient-heat-result}
		Comparison of 1D transient heat conduction results across a ﬁlm. $L$ = 10nm, $L$ = 100nm, $L$ = 1000nm.}
\end{figure}

\subsection{Heat transfer in the 2D square domain}
To validate the effectiveness of the UGKWP method and the IUGKP method for phonon transport in multidimensional physical space, this section investigates the heat conduction problem in a 2D square domain.
Specifically, the top boundary is maintained at a high temperature $T_H$, while the remaining boundaries are held at a lower temperature $T_C$ as illustrated in Fig.\ref{2d-square}.

\begin{figure}[htb]	\label{2d-square}
	\centering	
	\includegraphics[height=0.45\textwidth]{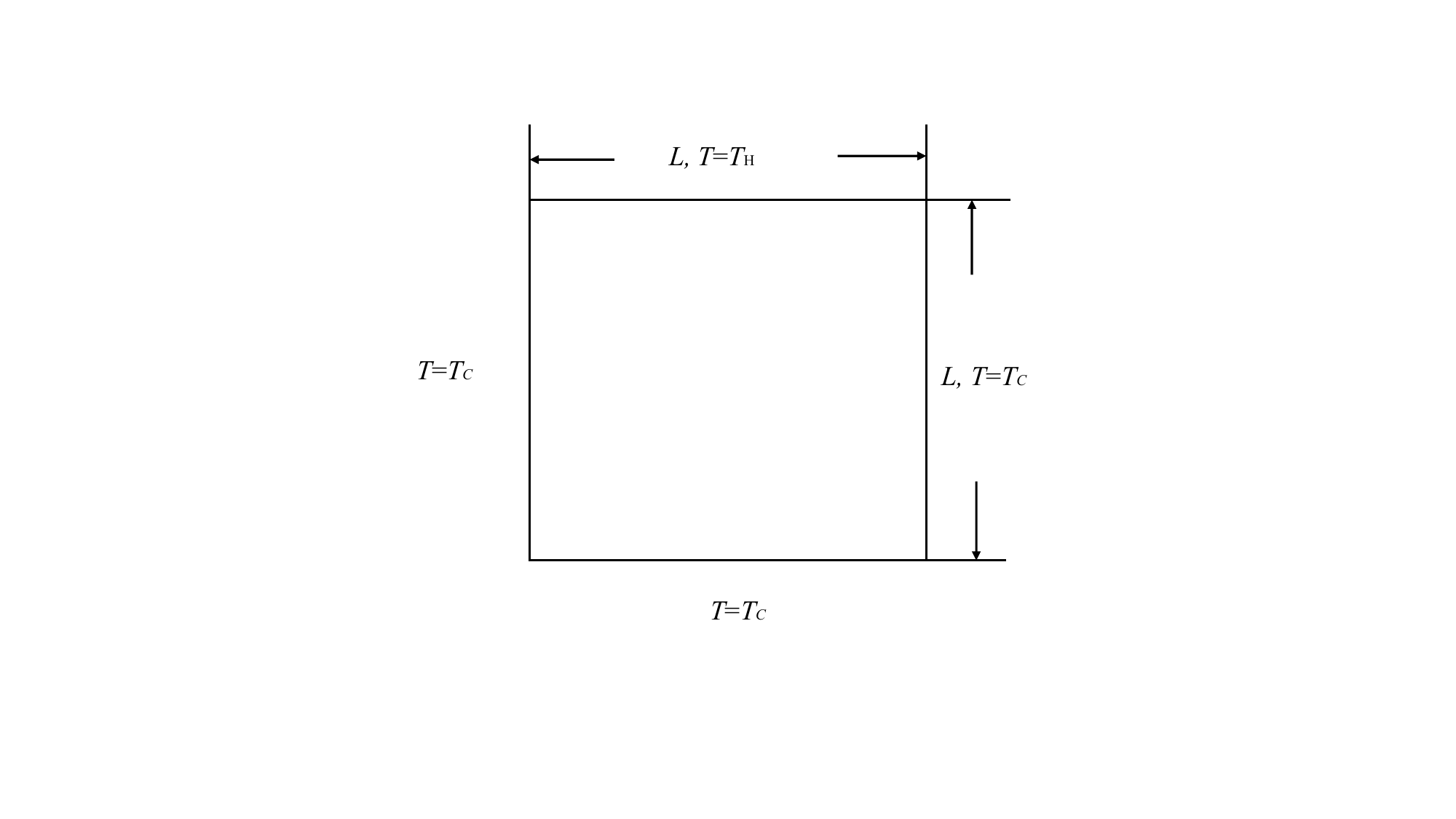}
	\caption{\label{2d-square}
		Computational domain and boundary conditions of heat transfer in the 2D square domain. }
\end{figure}

In this case, we employ a two-dimensional uniform grid, discretized into 40 equally spaced points in each direction, resulting in a total of 1,600 grid points.
In this case, $N_{ref}=300$ and $N_{min} = 20$ are used in each cell for $L=10$ nm, $L=100$ nm, and $L=1\mu$m to balancing computational efficiency and statistical noise.
Moreover, $N_B=20$ for each phonon branch. 
Furthermore, an additional 200 steps were incorporated into the two-dimensional simulation to reduce statistical noise through statistical averaging.
The computation results for different physical scales are shown in Fig.\ref{2d-square-result}. 
The black solid lines in the figure represent the contour lines obtained using the UGKWP method, while the white dashed lines represent those computed with the IUGKP method. 
As can be seen, the UGKWP method and the IUGKP method exhibits very good agreement with each other in both the diffusive region and the ballistic region.

\begin{figure}[htb]	\label{2d-square-result}
	\centering	
    \includegraphics[height=0.40\textwidth]{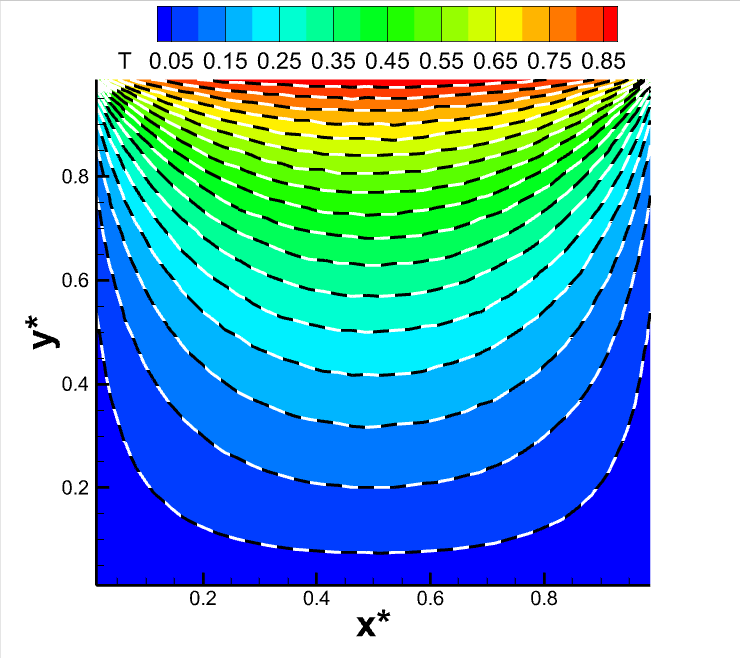}
	\includegraphics[height=0.40\textwidth]{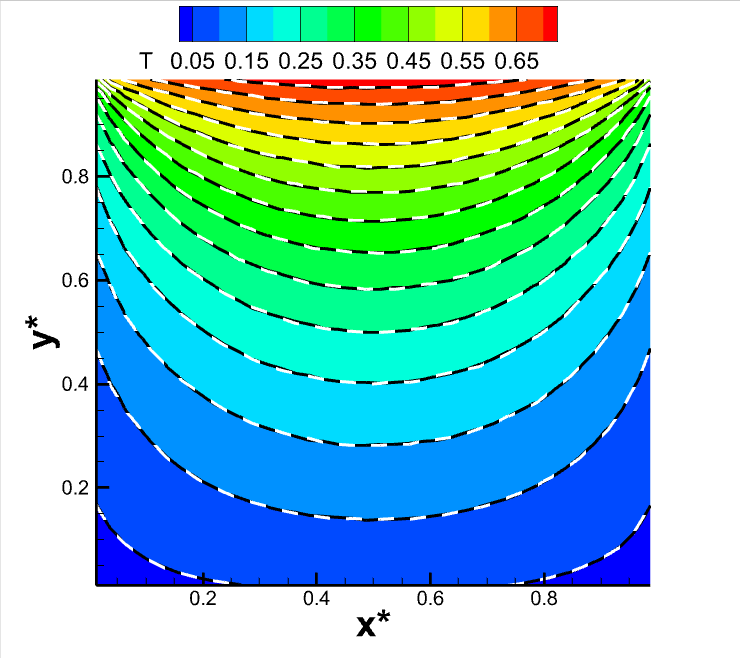}
    \includegraphics[height=0.40\textwidth]{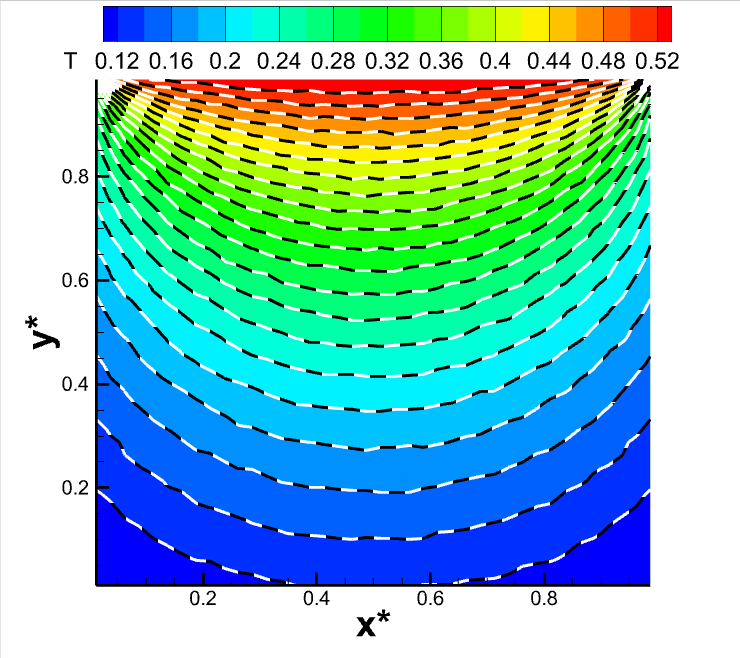}
	\caption{\label{2d-square-result}
		Comparison of 2D square heat transfer results, respectively corresponding: $L=10$ nm, $L=100$ nm, and $L=1\mu$m. }
\end{figure}

Table~.\ref{2d-iugkp-ugkwp} gives the total CPU time for the UGKWP method and the IUGKP method with 200 statistical averaging steps.
The IUGKP method achieves a two‐order‐of‐magnitude speed‐up in computational time upon convergence compared to the UGKWP method (the UGKWP method without frequency adaptive sampling).

In addition to the fast convergence characteristics of the IUGKP algorithm, the significant difference is also attributed to its use of adaptive frequency-space sampling, which significantly reduces the total number of particles in the IUGKP method. 
Table~.\ref{particle-number-comparison} shows the particle number for the three cases of the UGKWP (without frequency adaptive sampling) method and the IUGKP method.

At the same time, we compared the computational times of the UGKWP method, when frequency-space adaptive sampling is also applied, with those of the IUGKP method, as shown in Table~.\ref{2d-iugkp-ugkwp-all-adaptive}.
It can be seen that even when both methods employ adaptive sampling in the frequency space, the IUGKP method still achieves a computational speedup of several orders of magnitude.
This is because, after applying adaptive sampling in the frequency space, the number of sampled particles for both methods is almost the same. 
The speedup is attributed to the inherent fast-convergence properties of the steady-state IUGKP method and its simple data structure, which result in a lower computational cost per step.

\begin{table}[htp]
	\small
	\begin{center}
		\def\temptablewidth{1.0\textwidth}
		{\rule{\temptablewidth}{1pt}}
		\begin{tabular*}{\temptablewidth}{@{\extracolsep{\fill}}c|c|c|c}
			Method & 10nm &  100nm & 1000nm  \\
			\hline
			UGKWP & 23300s (2300 total steps) & 41450s (4100 total steps) & 34800s (5000 steps) \\ 	
			IUGKP & 62s (210 total steps) & 66s (220 total steps) & 94s (300 total steps)\\ 	
                CPU Time speedup & 376 & 628 & 370 \\ 	
		\end{tabular*}
		{\rule{\temptablewidth}{0.1pt}}
	\end{center}
	\vspace{-4mm} \caption{\label{2d-iugkp-ugkwp} Comparison of the Computational costs of two-dimensional heat conduction between the UGKWP method and the IUGKP method (the UGKWP method without frequency adaptive sampling).}
\end{table}

\begin{table}[htp]
	\small
	\begin{center}
		\def\temptablewidth{1.0\textwidth}
		{\rule{\temptablewidth}{1pt}}
		\begin{tabular*}{\temptablewidth}{@{\extracolsep{\fill}}c|c|c|c}
			Method & 10nm &  100nm & 1000nm  \\
			\hline
			UGKWP & 300 $\times$ 20 $\times$ 2 & 300 $\times$ 20 $\times$ 2 & 200 $\times$ 20 $\times$ 2 \\ 	
			IUGKP & 1062 & 1062 & 1062\\ 	
                Particle Number Ratio & 11.29 & 11.29 & 7.53 \\ 	
		\end{tabular*}
		{\rule{\temptablewidth}{0.1pt}}
	\end{center}
	\vspace{-4mm} \caption{\label{particle-number-comparison} Comparison of the particle number of two-dimensional heat conduction between the UGKWP method and the IUGKP method (the UGKWP method without frequency adaptive sampling).}
\end{table}

\begin{table}[htp]
	\small
	\begin{center}
		\def\temptablewidth{1.0\textwidth}
		{\rule{\temptablewidth}{1pt}}
		\begin{tabular*}{\temptablewidth}{@{\extracolsep{\fill}}c|c|c|c}
			Method & 10nm &  100nm & 1000nm  \\
			\hline
			UGKWP & 1960s (2300 total steps) & 3769s (4100 total steps) & 4580s (5000 steps) \\ 	
			IUGKP & 62s (210 total steps) & 66s (220 total steps) & 94s (300 total steps)\\ 	
                CPU Time speedup & 32 & 57 & 48 \\ 	
		\end{tabular*}
		{\rule{\temptablewidth}{0.1pt}}
	\end{center}
	\vspace{-4mm} \caption{\label{2d-iugkp-ugkwp-all-adaptive} Comparison of the Computational costs of two-dimensional heat conduction between the UGKWP method and the IUGKP method (the UGKWP method using frequency adaptive sampling).}
\end{table}

\subsection{Heat transfer in the 3D cubic domain}
To demonstrate the generality of the proposed algorithm, this section simulates the heat conduction problem in a three-dimensional cubic cavity. Computational domain and boundary conditions are shown in Fig~.\ref{3d}
\begin{figure}[htb]	\label{3d}
	\centering	
	\includegraphics[height=0.50\textwidth]{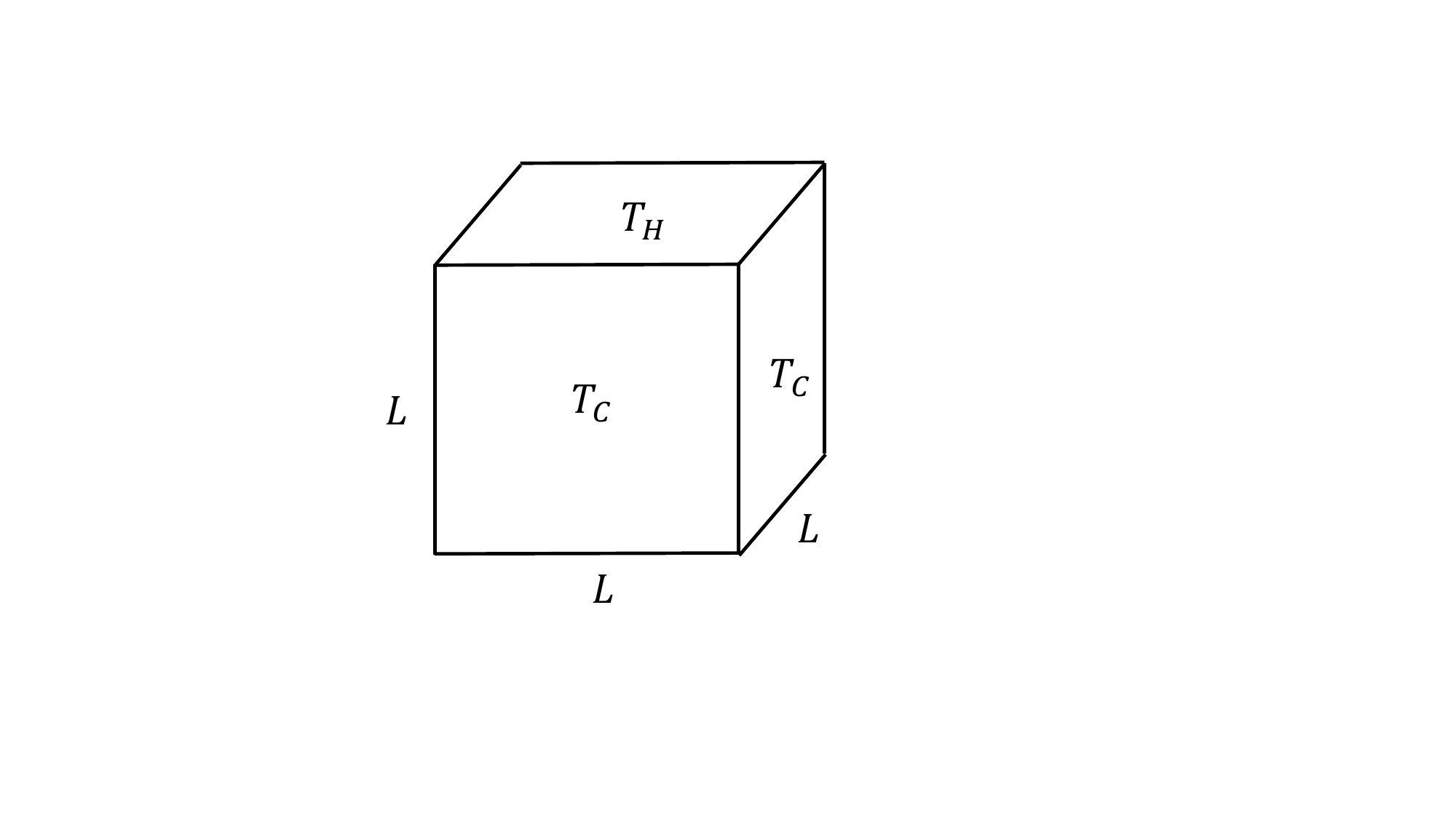}
	\caption{\label{3d}
		Computational domain and boundary conditions of Heat transfer in the 3D cubic domain. }
\end{figure}
All boundaries are isothermal boundaries, with the top boundary held at temperature $T_H$ and the remaining boundaries at temperature $T_c$.

In this case, $N_{ref}=200$ and $N_{min} = 20$ are used in each cell.
Moreover, $N_B=20$ for each phonon branch. 
Furthermore, an additional 200 steps were incorporated into the three-dimensional simulation to reduce statistical noise through statistical averaging.
The computation results of IUGKP for $L = 100$ nm are shown in Fig~.\ref{3d-cubic-result}. 
\begin{figure}[htb]	\label{3d-cubic-result}
	\centering	
    \includegraphics[height=0.40\textwidth]{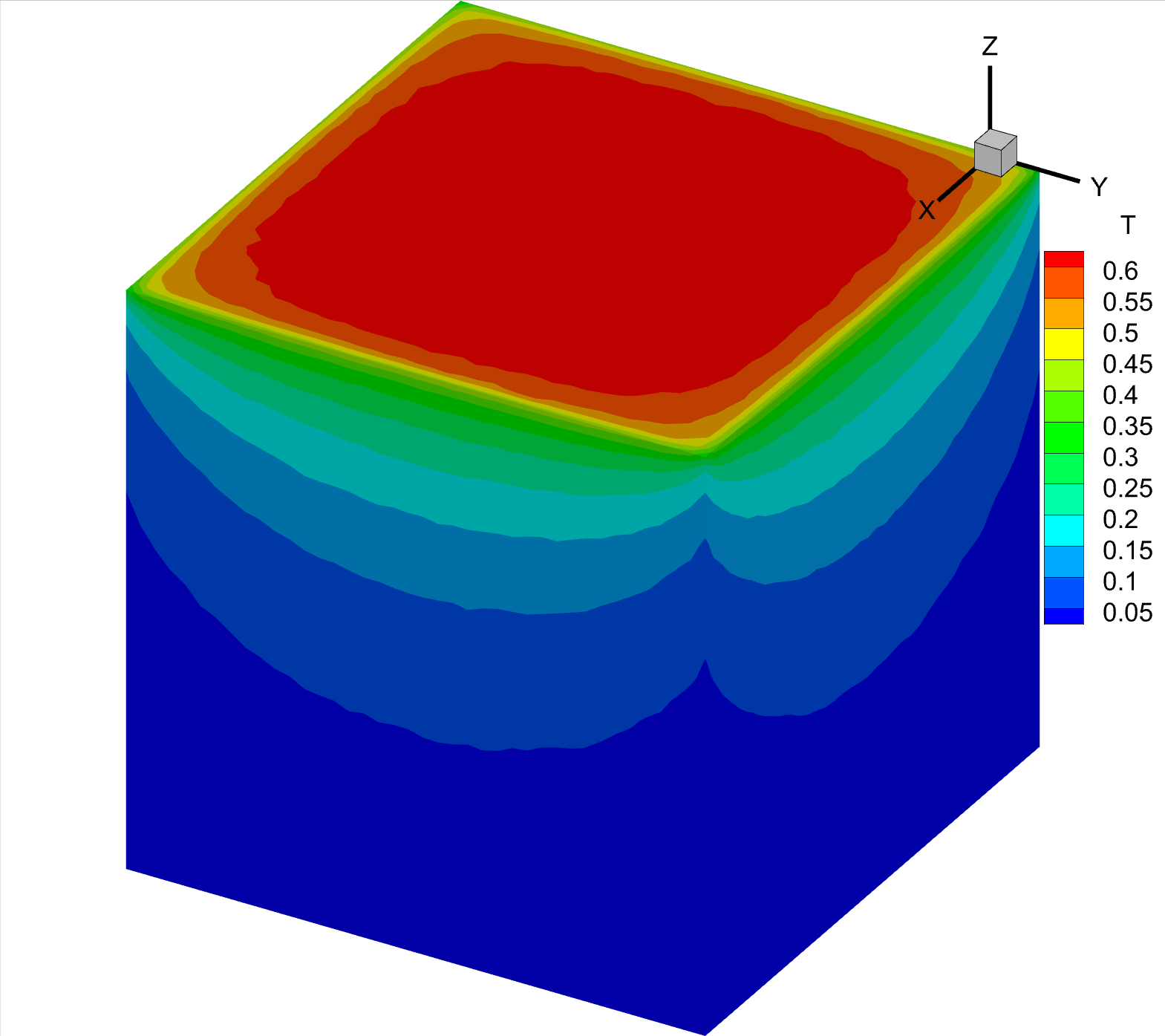}
        \includegraphics[height=0.40\textwidth]{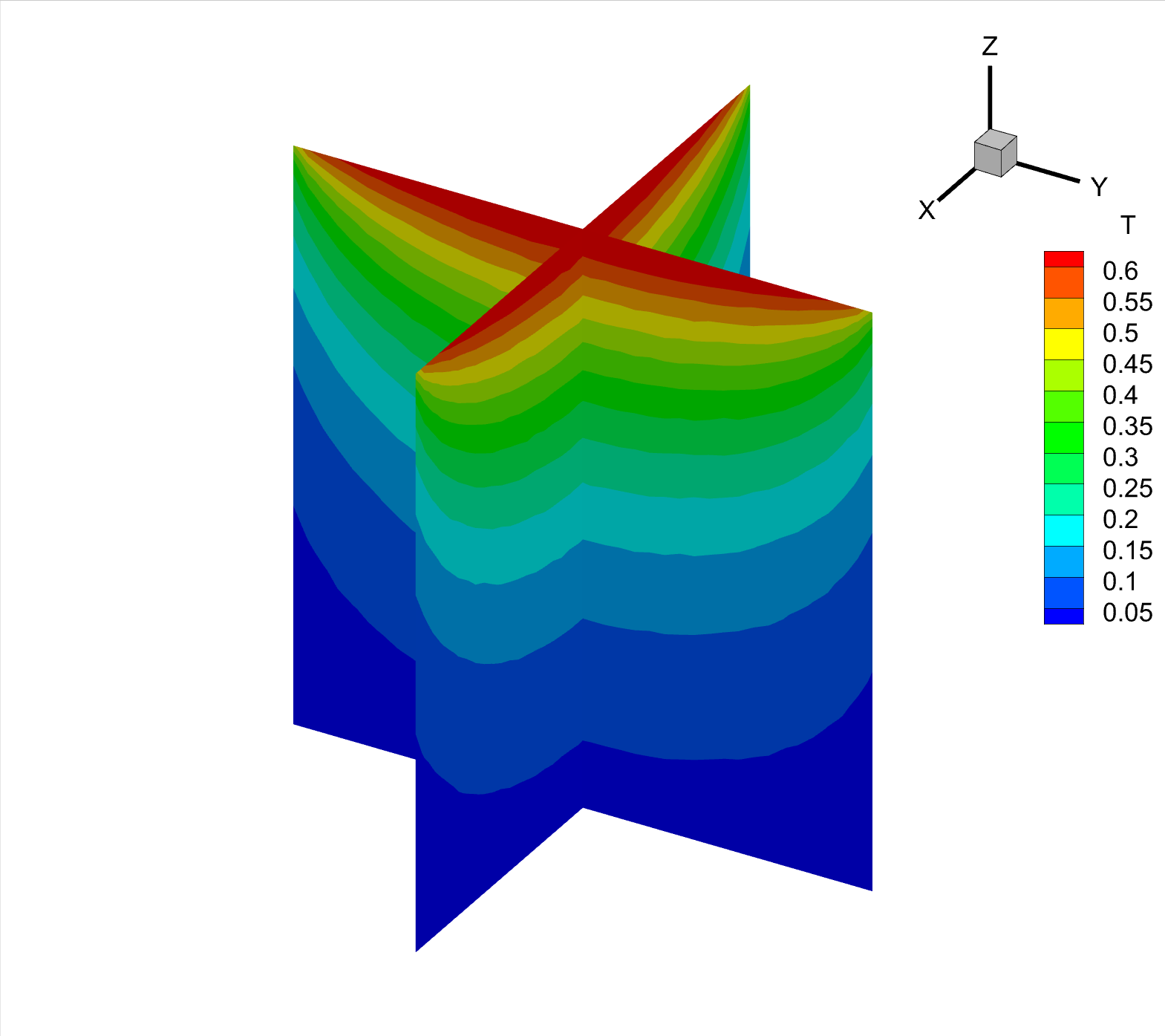}
	\caption{\label{3d-cubic-result}
		 Contour of the heat conduction in the 3D cubic. Left: 3D contour; Right: slice of $x^*=0.5$ plane and slice of $y^*=0.5$ plane.}
\end{figure}
The quantitative results, as shown in Fig~.\ref{3d-cubic-line}, clearly demonstrate that the IUGKP method agrees very well with the reference solution \cite{zhang2021fast}.
\begin{figure}[htb]	\label{3d-cubic-line}
	\centering	
    \includegraphics[height=0.40\textwidth]{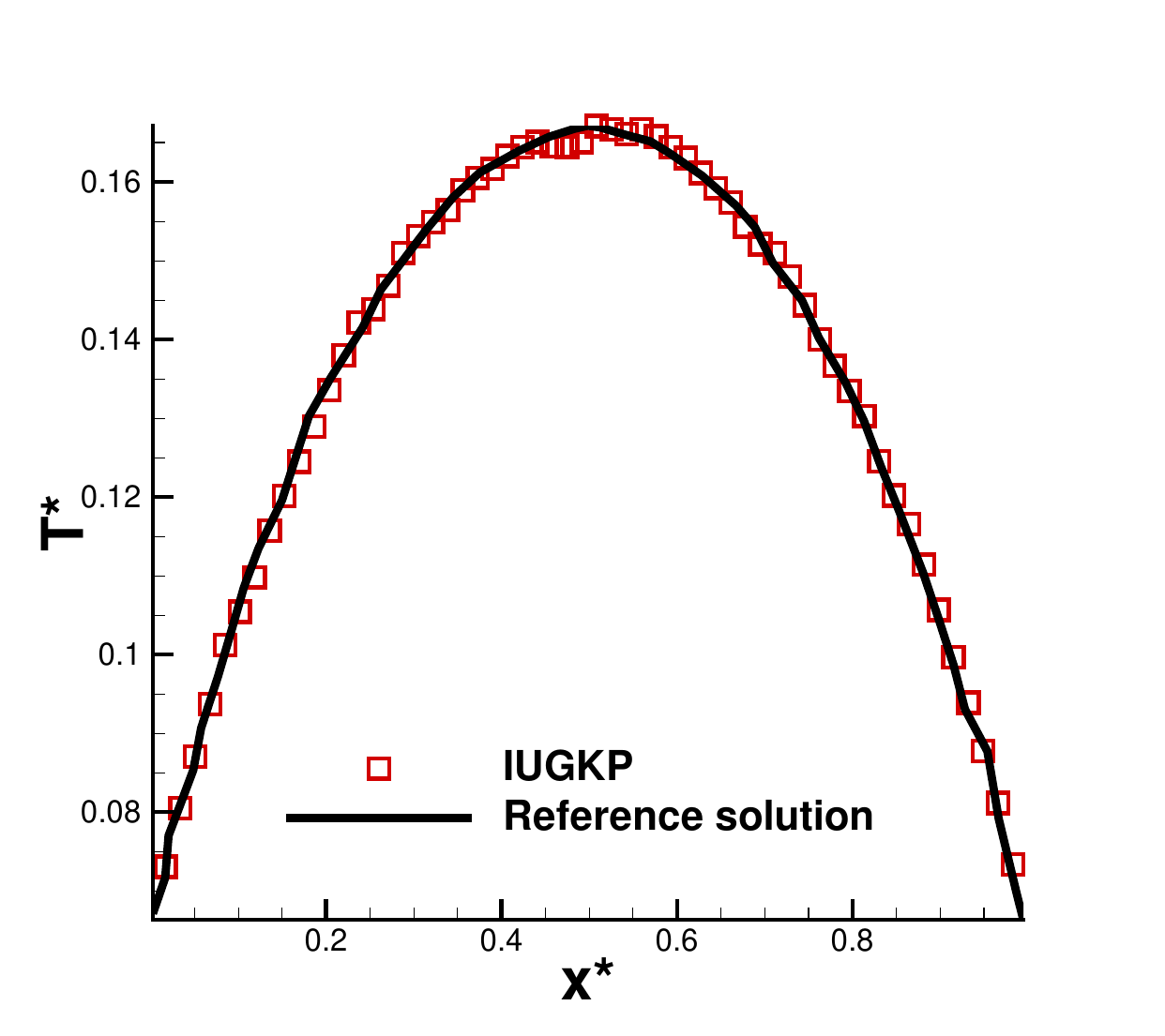}
        \includegraphics[height=0.40\textwidth]{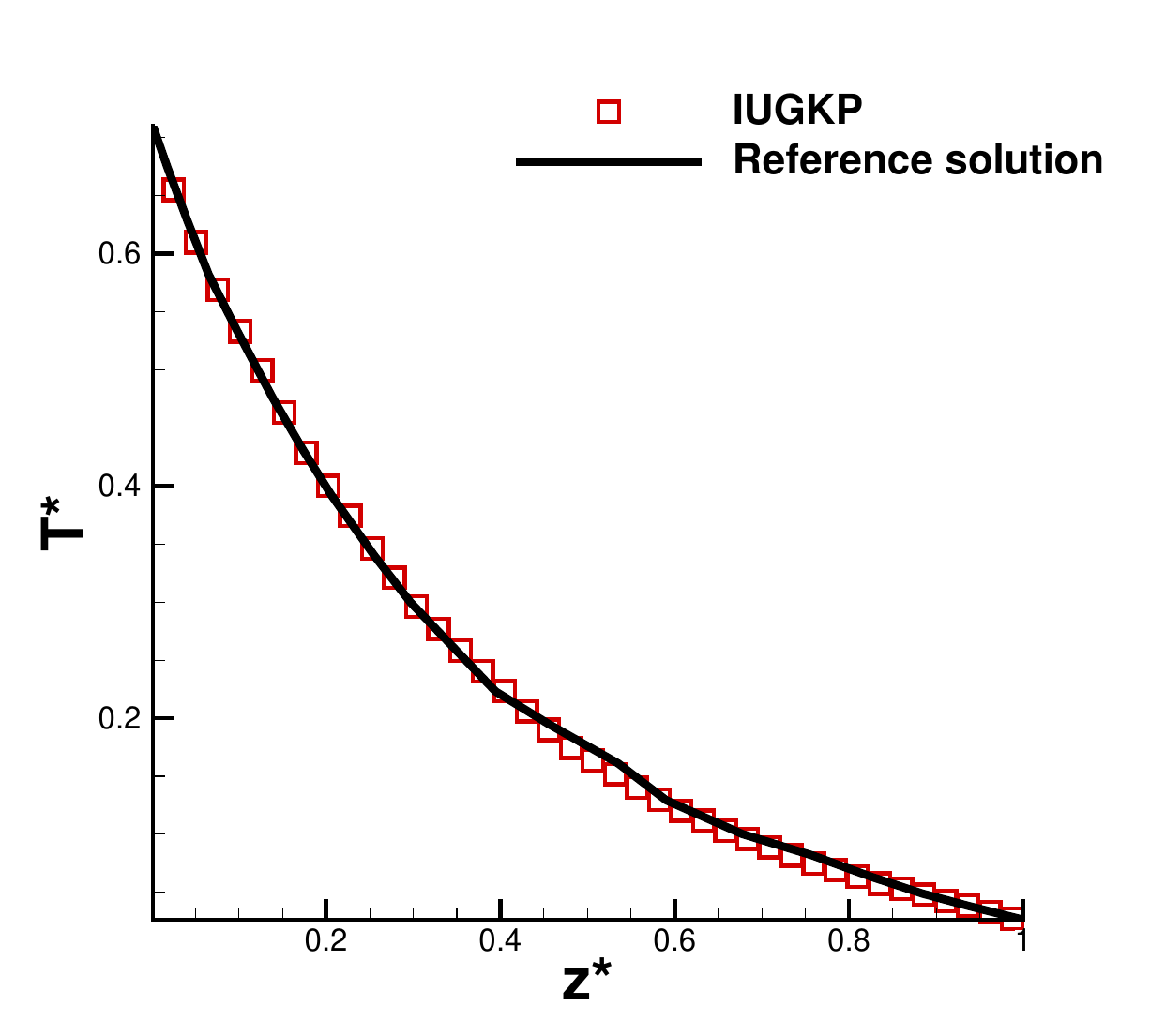}
	\caption{\label{3d-cubic-line}
		 Quantitative results comparison. Left: Temperature distribution from [0, 0.5, 0.5] to [1, 0.5, 0.5]; Right: Temperature distribution from [0.5, 0.5, 1] to [0.5, 0.5, 0].}
\end{figure}
In terms of computational efficiency, this test case executed a total of 240 steps, including 40 computation steps and 200 averaging steps, with a total runtime of 40 minutes.
This test case consumed 2.0GB of runtime memory.
It is nearly impossible to solve the six-dimensional BTE using the DOM method on a single CPU core in a personal laptop. This highlights the significant advantages of the IUGKP method in terms of both memory usage and computational speed.

\subsection{Heat dissipation simulation in three-dimensional structures}
To further validate the IUGKP method's capability in capturing multi-scale non-equilibrium phonon transport in multi-dimensional problems, this section presents a heat dissipation simulation in three-dimensional structures, shown as Fig~.\ref{heat-dissipation-domain}. 
The computational domain has dimensions $L_x$, $L_y$, and $L_z$ for length, width, and height, respectively. 
At the center of the top surface, there is a square high-temperature heat source with a side length of $L_h$ and a temperature of $T_h$. 
At the center of the bottom surface, there is a rectangular low-temperature heat sink with a temperature of $T_c$, where its length and width are $L_c$ and $L_y$, respectively.
The high-temperature heat source and the low-temperature heat sink are maintained under isothermal boundary conditions, while all other walls are subject to adiabatic boundary conditions.
\begin{figure}[htb]	\label{heat-dissipation-domain}
	\centering	
    \includegraphics[height=0.40\textwidth]{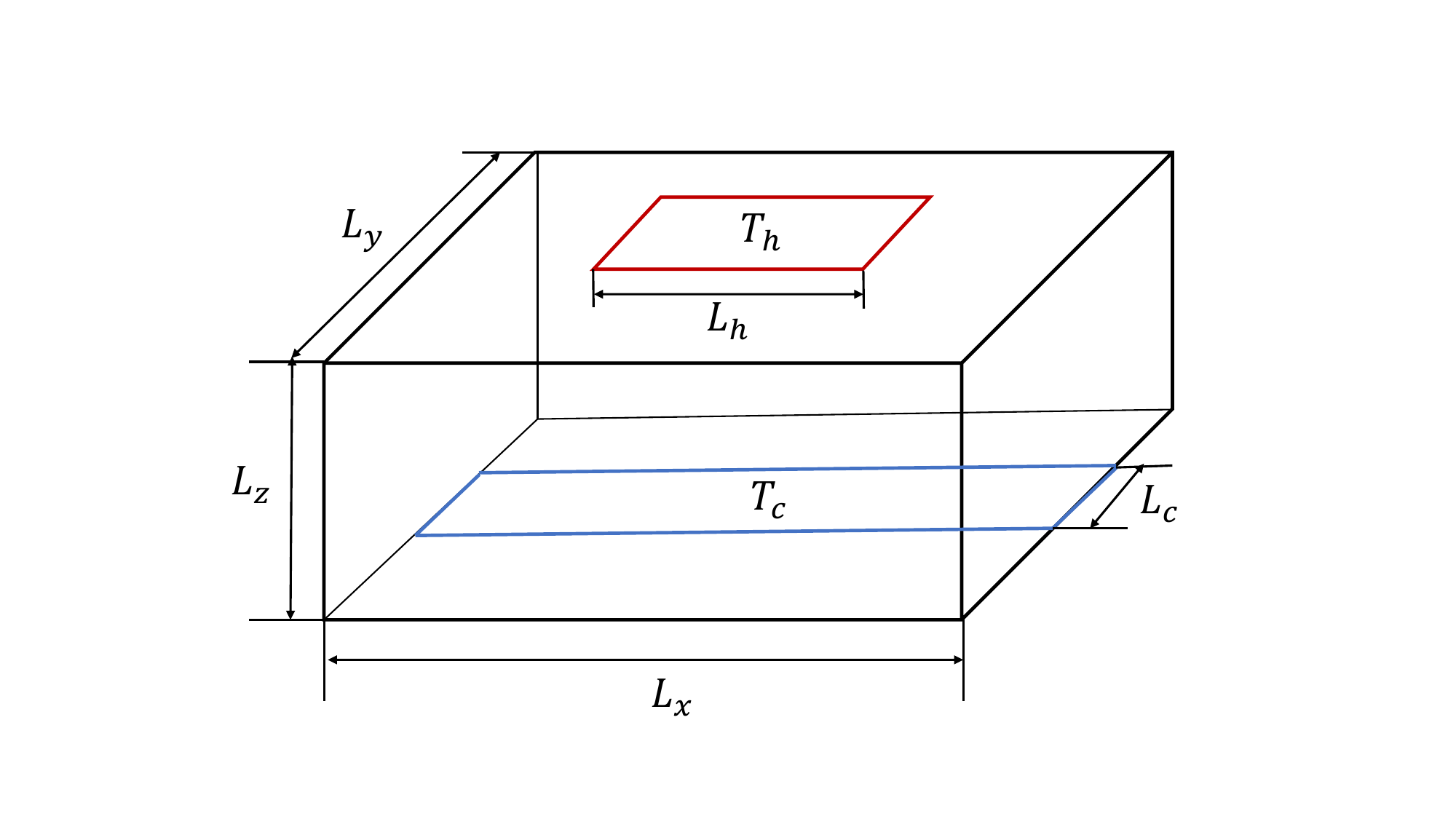}
	\caption{\label{heat-dissipation-domain}
		Computational domain and boundary conditions for Heat dissipation simulation in three-dimensional structures\cite{zhang2021fast}.}
\end{figure}
Moreover, $L_h=L_c=\frac{1}{4}L_x$, $L_x=L_y=2L_z=0.1\mu$m.
In this case, $N_{ref}=200$ and $N_{min} = 20$ are used in each cell.
Moreover, $N_B=20$ for each phonon branch. 
Furthermore, an additional 200 steps were incorporated into the three-dimensional simulation to reduce statistical noise through statistical averaging.
The computation result is shown in Fig~.\ref{3d-cubic-result}. 
\begin{figure}[htb]	\label{3d-cubic-result}
	\centering	
    \includegraphics[height=0.40\textwidth]{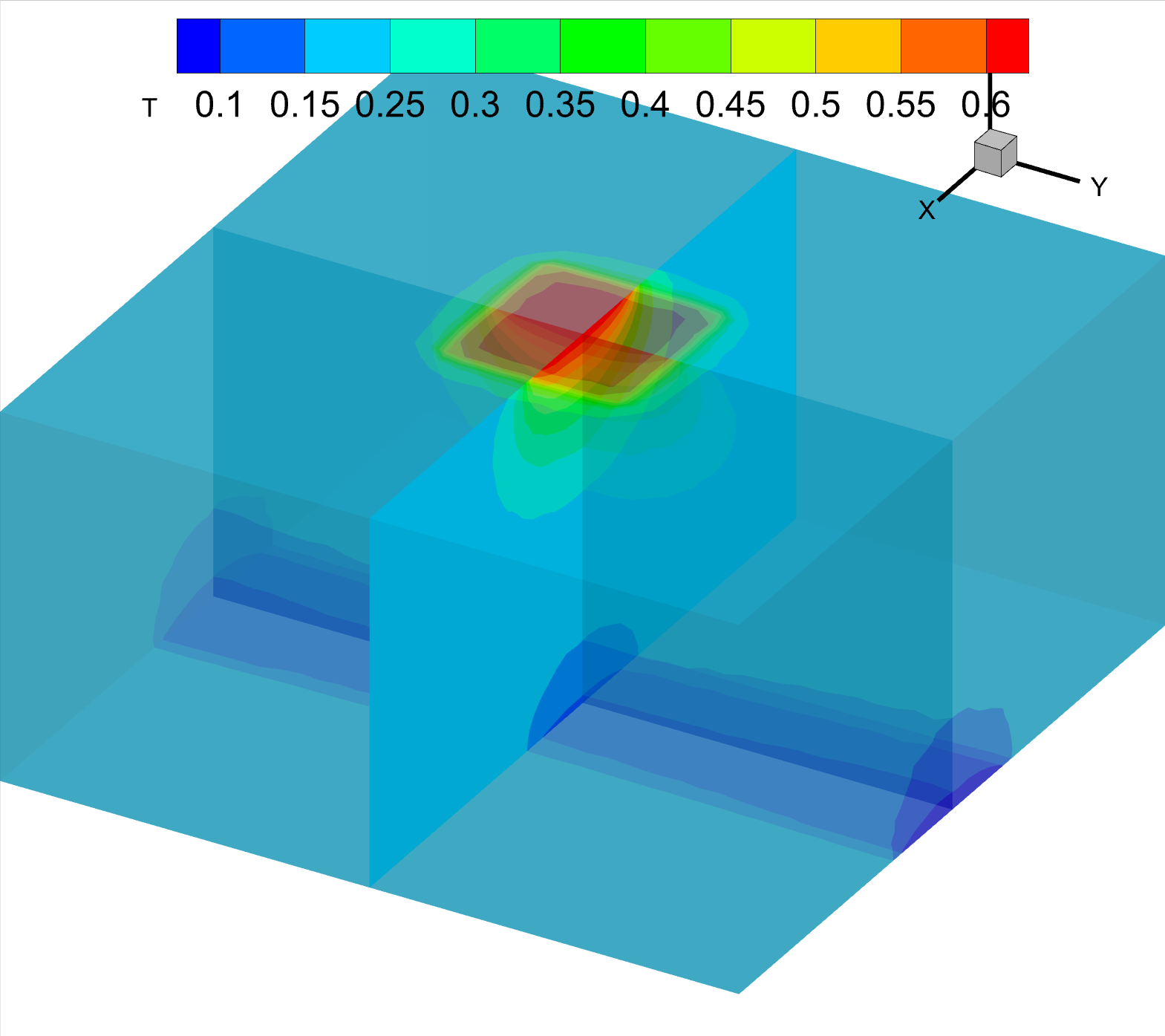}
    \includegraphics[height=0.40\textwidth]{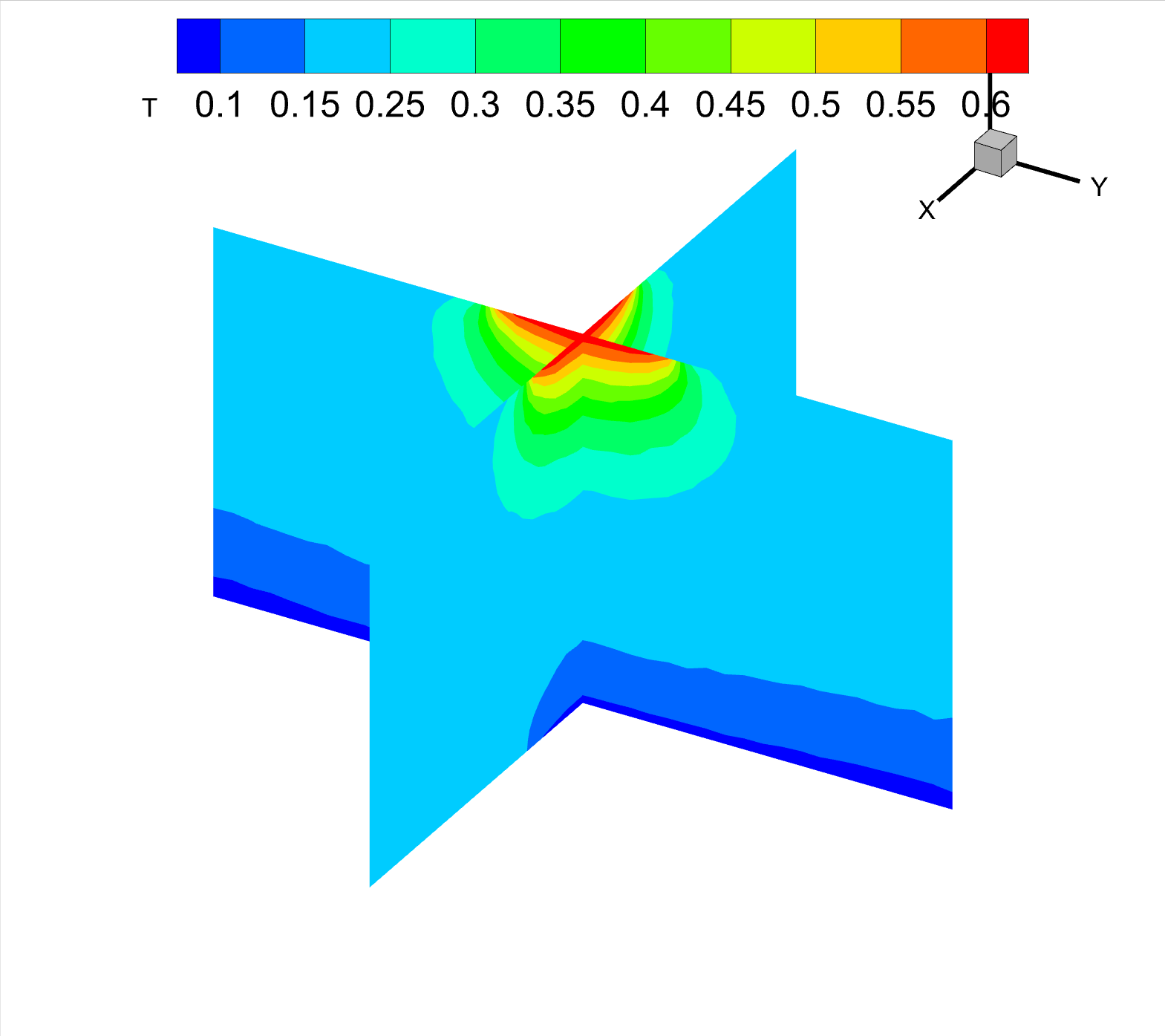}
	\caption{\label{3d-cubic-result}
		 Contour of the heat dissipation simulation in three-dimensional structures. Left: 3D contour; Right: slice of $x^*=0.5$ plane and slice of $y^*=0.5$ plane.}
\end{figure}
The quantitative results, as shown in Fig~.\ref{3d-un-line}, clearly demonstrate that the IUGKP method agrees very well with the reference solution \cite{zhang2021fast}.
\begin{figure}[htb]	\label{3d-un-line}
	\centering	
    \includegraphics[height=0.40\textwidth]{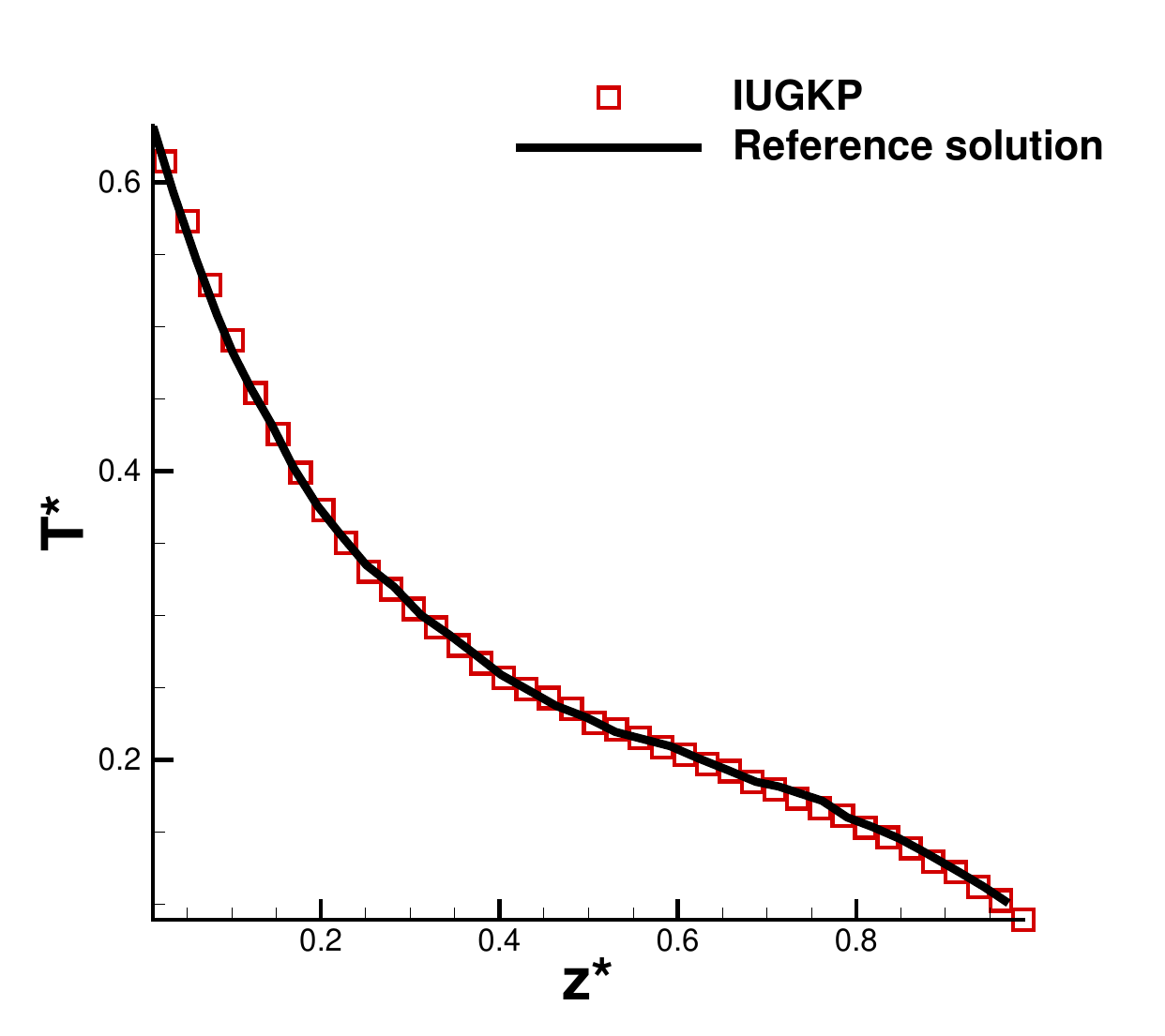}
	\caption{\label{3d-un-line}
		 Quantitative results comparison, temperature distribution from [0.5, 0.5, 1.0] to [0.5, 0.5, 0].}
\end{figure}

In terms of computational efficiency, this test case executed a total of 280 steps, including 80 computation steps and 200 averaging steps, with a total runtime of 100 minutes.
This test case consumed 2.0 GB of runtime memory.
This means that the IUGKP algorithm proposed in this paper is already capable of simulating large-scale, three-dimensional, multiscale heat conduction.

\section{Conclusion}

This paper introduces two novel methods—the UGKWP and IUGKP—for solving the phonon Boltzmann equation with consideration of phonon polarization and dispersion relations. Both methods demonstrate remarkable adaptability across various heat conduction regimes. The UGKWP method naturally transitions between different transport regimes: in the diffusive limit, it recovers Fourier's law as particle contributions become negligible, while in the ballistic limit, non-equilibrium flux is completely characterized by free-streaming particles.
To enhance computational efficiency for steady-state multi-frequency phonon transport problems, we developed the IUGKP method. The IUGKP method achieves rapid convergence through different mechanisms depending on the regime: the macroscopic prediction equation dominates in the diffusive limit, while large-step particle transport accelerates convergence in the ballistic regime.
While the explicit time-marching UGKWP method converges more slowly than IUGKP due to its unsteady BGK equation foundation, it shows considerable promise for unsteady computations. Extensive numerical testing across diffusive to ballistic regimes validates both methods' accuracy and demonstrates their exceptional multi-scale adaptability, effectively bridging different transport regimes.
Additionally, we developed an adaptive particle sampling method in frequency space.
This innovation reduces particle sampling requirements to the same order of magnitude as the gray model for both the UGKWP and the IUGKP method.
A notable achievement of the IUGKP method is its computational efficiency: large-scale three-dimensional, multi-scale heat conduction simulations can be completed within 1-2 hours on a single-core personal laptop. This performance demonstrates significant advantages in both memory usage and convergence speed compared to traditional DOM-type deterministic methods.
Future work will focus on extending these methods to solve multi-scale phonon transport problems in systems with arbitrary temperature differences, further expanding their practical applications in thermal transport analysis.


\section{Acknowledgements}

The current research is supported by National Key R\&D Program of China (Grant Nos. 2022YFA1004500), National Science Foundation of China (12172316, 92371107), and Hong Kong research grant council (16301222, 16208324).

\bibliographystyle{unsrt}
\bibliography{jixingbib}

\begin{thebibliography}{10}

\bibitem{TCAD_application_intel_2021_review}
Mark~A. Stettler, Stephen~M. Cea, Sayed Hasan, Lei Jiang, Patrick~H. Keys, Colin~D. Landon, Prabakar Marepalli, Daniel Pantuso, and Cory~E. Weber.
\newblock Industrial {TCAD}: Modeling atoms to chips.
\newblock {\em IEEE Transactions on Electron Devices}, 68(11):5350--5357, 2021.

\bibitem{murthy2005review}
Jayathi~Y Murthy, Sreekant~VJ Narumanchi, Pascual-Gutierrez Jose'A, Tianjiao Wang, Chunjian Ni, and Sanjay~R Mathur.
\newblock Review of multiscale simulation in submicron heat transfer.
\newblock {\em International Journal for Multiscale Computational Engineering}, 3(1), 2005.

\bibitem{review_2019_CMOS}
C.~Prasad.
\newblock A review of self-heating effects in advanced {CMOS} technologies.
\newblock {\em IEEE Transactions on Electron Devices}, 66(11):4546--4555, 2019.

\bibitem{HUA2023chapter}
Yu-Chao Hua, Yang Shen, Zheng-Lai Tang, Dao-Sheng Tang, Xin Ran, and Bing-Yang Cao.
\newblock Chapter eight - near-junction thermal managements of electronics.
\newblock {\em Advances in Heat Transfer}, 56:355--434, 2023.

\bibitem{phononsnanoscale}
Parallel Treatment Of Electrons~Molecules Phonons.
\newblock Nanoscale energy transport and conversion a parallel treatment of electrons molecules phonons and photons mit pappalardo series in mechanical engineering by gang chen.

\bibitem{RevModPhys.90.041002}
Xiaokun Gu, Yujie Wei, Xiaobo Yin, Baowen Li, and Ronggui Yang.
\newblock Colloquium: phononic thermal properties of two-dimensional materials.
\newblock {\em Rev. Mod. Phys.}, 90:041002, Nov 2018.

\bibitem{RevModPhys.94.025002}
Jie Chen, Xiangfan Xu, Jun Zhou, and Baowen Li.
\newblock Interfacial thermal resistance: Past, present, and future.
\newblock {\em Rev. Mod. Phys.}, 94:025002, Apr 2022.

\bibitem{ziman2001electrons}
John~M Ziman.
\newblock {\em Electrons and phonons: the theory of transport phenomena in solids}.
\newblock Oxford university press, 2001.

\bibitem{chattopadhyay2014comparative}
Ankur Chattopadhyay and Arvind Pattamatta.
\newblock A comparative study of submicron phonon transport using the boltzmann transport equation and the lattice boltzmann method.
\newblock {\em Numerical Heat Transfer, Part B: Fundamentals}, 66(4):360--379, 2014.

\bibitem{barry2022boltzmann}
Matthew~C Barry, Nitish Kumar, and Satish Kumar.
\newblock {Boltzmann} transport equation for thermal transport in electronic materials and devices.
\newblock {\em Annual Review of Heat Transfer}, 24:131--172, 2022.

\bibitem{mazumder_boltzmann_2022}
Sandip Mazumder.
\newblock {Boltzmann} transport equation based modeling of phonon heat conduction: {Progress} and challenges.
\newblock {\em Annual Review of Heat Transfer}, 24:71--130, 2022.

\bibitem{guo_progress_DUGKS}
Zhaoli Guo and Kun Xu.
\newblock Progress of discrete unified gas-kinetic scheme for multiscale flows.
\newblock {\em Adva. Aerodyn.}, 3(1):6, Feb 2021.

\bibitem{peraud_monte_2014}
Jean-Philippe~M. Peraud, Colin~D. Landon, and Nicolas~G. Hadjiconstantinou.
\newblock Monte carlo methods for solving the {B}oltzmann equation.
\newblock {\em Annu. Rev. Heat Transfer}, 17:205--265, 2014.

\bibitem{lacroix2005monte}
David Lacroix, Karl Joulain, and Denis Lemonnier.
\newblock Monte {Carlo} transient phonon transport in silicon and germanium at nanoscales.
\newblock {\em Physical Review B—Condensed Matter and Materials Physics}, 72(6):064305, 2005.

\bibitem{GuoZl13LB}
Zhaoli Guo and Chang Shu.
\newblock {\em Lattice {B}oltzmann method and its applications in engineering}, volume~3.
\newblock World Scientific, 2013.

\bibitem{stamnes1988dom}
Knut Stamnes, S-Chee Tsay, Warren Wiscombe, and Kolf Jayaweera.
\newblock Numerically stable algorithm for discrete-ordinate-method radiative transfer in multiple scattering and emitting layered media.
\newblock {\em Applied optics}, 27(12):2502--2509, 1988.

\bibitem{murthy1998domunstruct}
Jayathi~Y Murthy and SR~Mathur.
\newblock Finite volume method for radiative heat transfer using unstructured meshes.
\newblock {\em Journal of thermophysics and heat transfer}, 12(3):313--321, 1998.

\bibitem{SyedAA14LargeScale}
Syed~Ashraf Ali, Gautham Kollu, Sandip Mazumder, P.~Sadayappan, and Arpit Mittal.
\newblock Large-scale parallel computation of the phonon {B}oltzmann transport equation.
\newblock {\em Int. J. Therm. Sci}, 86:341 -- 351, 2014.

\bibitem{FivelandVA96Acceleration}
V.~A. Fiveland and J.~P. Jessee.
\newblock Acceleration schemes for the discrete ordinates method.
\newblock {\em J. Thermophys. Heat Transfer.}, 10(3):445--451, 1996.

\bibitem{guo2016dugksphonon}
Zhaoli Guo and Kun Xu.
\newblock Discrete unified gas kinetic scheme for multiscale heat transfer based on the phonon boltzmann transport equation.
\newblock {\em International Journal of Heat and Mass Transfer}, 102:944--958, 2016.

\bibitem{luo2017dugksphonon}
Xiao-Ping Luo and Hong-Liang Yi.
\newblock A discrete unified gas kinetic scheme for phonon boltzmann transport equation accounting for phonon dispersion and polarization.
\newblock {\em International Journal of Heat and Mass Transfer}, 114:970--980, 2017.

\bibitem{zhang2019dugksphonon}
Chuang Zhang and Zhaoli Guo.
\newblock Discrete unified gas kinetic scheme for multiscale heat transfer with arbitrary temperature difference.
\newblock {\em International Journal of Heat and Mass Transfer}, 134:1127--1136, 2019.

\bibitem{Chuang17gray}
Chuang Zhang, Zhaoli Guo, and Songze Chen.
\newblock Unified implicit kinetic scheme for steady multiscale heat transfer based on the phonon {B}oltzmann transport equation.
\newblock {\em Phys. Rev. E}, 96:063311, Dec 2017.

\bibitem{zhang2023acceleration}
Chuang Zhang, Samuel Huberman, Xinliang Song, Jin Zhao, Songze Chen, and Lei Wu.
\newblock Acceleration strategy of source iteration method for the stationary phonon boltzmann transport equation.
\newblock {\em International Journal of Heat and Mass Transfer}, 217:124715, 2023.

\bibitem{zhang2025synthetic}
Chuang Zhang, Qin Lou, and Hong Liang.
\newblock Synthetic iterative scheme for thermal applications in hotspot systems with large temperature variance.
\newblock {\em International Journal of Heat and Mass Transfer}, 236:126374, 2025.

\bibitem{qian1992lattice}
Yue-Hong Qian, Dominique d'Humi{\`e}res, and Pierre Lallemand.
\newblock Lattice bgk models for navier-stokes equation.
\newblock {\em EPL (Europhysics Letters)}, 17(6):479, 1992.

\bibitem{DSMC_book_1994}
G~A Bird.
\newblock {\em {Molecular Gas Dynamics And The Direct Simulation Of Gas Flows}}.
\newblock Oxford University Press, 05 1994.

\bibitem{DSMC_phonon_1994}
R.~B. Peterson.
\newblock {Direct Simulation of Phonon-Mediated Heat Transfer in a Debye Crystal}.
\newblock {\em Journal of Heat Transfer}, 116(4):815--822, 11 1994.

\bibitem{mazumder2001monte}
Sandip Mazumder and Arunava Majumdar.
\newblock Monte {Carlo} study of phonon transport in solid thin films including dispersion and polarization.
\newblock {\em J. Heat Transfer}, 123(4):749--759, 2001.

\bibitem{mittal2010monte}
Arpit Mittal and Sandip Mazumder.
\newblock Monte carlo study of phonon heat conduction in silicon thin films including contributions of optical phonons.
\newblock {\em J. Heat Transfer}, 132(5):052402, 2010.

\bibitem{PJP11MC}
Jean-Philippe~M. P\'eraud and Nicolas~G. Hadjiconstantinou.
\newblock Efficient simulation of multidimensional phonon transport using energy-based variance-reduced monte carlo formulations.
\newblock {\em Phys. Rev. B}, 84:205331, Nov 2011.

\bibitem{peraud2015derivationmonte}
Jean-Philippe~M P{\'e}raud and Nicolas~G Hadjiconstantinou.
\newblock Adjoint-based deviational {Monte} {Carlo} methods for phonon transport calculations.
\newblock {\em Physical Review B}, 91(23):235321, 2015.

\bibitem{SILVA2024108954}
B.H. Silva, D.~Lacroix, M.~Isaiev, and L.~Chaput.
\newblock Monte carlo simulation of phonon transport from ab-initio data with nano-k.
\newblock {\em Computer Physics Communications}, 294:108954, 2024.

\bibitem{PATHAK2021108003}
Abhishek Pathak, Avinash Pawnday, Aditya~Prasad Roy, Amjad~J. Aref, Gary~F. Dargush, and Dipanshu Bansal.
\newblock {MCBTE}: A variance-reduced monte carlo solution of the linearized boltzmann transport equation for phonons.
\newblock {\em Comput. Phys. Commun.}, 265:108003, 2021.

\bibitem{zhu2019ugkwp}
Yajun Zhu, Chang Liu, Chengwen Zhong, and Kun Xu.
\newblock Unified gas-kinetic wave-particle methods. ii. multiscale simulation on unstructured mesh.
\newblock {\em Physics of Fluids}, 31(6), 2019.

\bibitem{liu2021ugkwp}
Chang Liu and Kun Xu.
\newblock Unified gas-kinetic wave-particle methods iv: multi-species gas mixture and plasma transport.
\newblock {\em Advances in Aerodynamics}, 3:1--31, 2021.

\bibitem{pu2025ugkwp}
Zhigang Pu and Kun Xu.
\newblock Unified gas-kinetic wave-particle method for multiscale flow simulation of partially ionized plasma.
\newblock {\em Journal of Computational Physics}, page 113918, 2025.

\bibitem{yang2024solid}
Xiaojian Yang, Wei Shyy, and Kun Xu.
\newblock Unified gas kinetic wave particle method for polydisperse gas-solid particle multiphase flow.
\newblock {\em Journal of Fluid Mechanics}, 983:A37, 2024.

\bibitem{yang2025rad}
Xiaojian Yang, Yajun Zhu, Chang Liu, and Kun Xu.
\newblock Unified gas-kinetic wave-particle method for frequency-dependent radiation transport equation.
\newblock {\em Journal of Computational Physics}, 522:113587, 2025.

\bibitem{yang2025turb}
Xiaojian Yang and Kun Xu.
\newblock {Wave}-{Particle} {Based} {Multiscale} {Modeling} and {Simulation} of {Non-equilibrium} {Turbulent} {Flows}.
\newblock {\em arXiv preprint arXiv:2503.07207}, 2025.

\bibitem{liu2025unified}
Hongyu Liu, Xiaojian Yang, Chuang Zhang, Xing Ji, and Kun Xu.
\newblock Unified gas-kinetic wave-particle method for multi-scale phonon transport.
\newblock {\em arXiv preprint arXiv:2505.09297}, 2025.

\bibitem{kaviany_2008}
Massoud Kaviany.
\newblock {\em Heat transfer physics}.
\newblock Cambridge University Press, 2008.

\bibitem{BGK}
Prabhu~Lal Bhatnagar, Eugene~P Gross, and Max Krook.
\newblock A model for collision processes in gases. {I}. {Small} amplitude processes in charged and neutral one-component systems.
\newblock {\em Physical Review}, 94(3):511, 1954.

\bibitem{liu2025implicit}
Hongyu Liu, Xiaojian Yang, Chuang Zhang, Xing Ji, and Kun Xu.
\newblock Implicit unified gas kinetic particle method for steady-state solution of multiscale phonon transport.
\newblock {\em arXiv preprint arXiv:2506.09465}, 2025.

\bibitem{ADAMS02fastiterative}
Marvin~L. Adams and Edward~W. Larsen.
\newblock Fast iterative methods for discrete-ordinates particle transport calculations.
\newblock {\em Prog. Nucl. Energ.}, 40(1):3 -- 159, 2002.

\bibitem{zhang2019implicit}
Chuang Zhang, Zhaoli Guo, and Songze Chen.
\newblock An implicit kinetic scheme for multiscale heat transfer problem accounting for phonon dispersion and polarization.
\newblock {\em International Journal of Heat and Mass Transfer}, 130:1366--1376, 2019.

\bibitem{terris2009modeling}
Damian Terris, Karl Joulain, Denis Lemonnier, and David Lacroix.
\newblock Modeling semiconductor nanostructures thermal properties: The dispersion role.
\newblock {\em Journal of Applied Physics}, 105(7), 2009.

\bibitem{zhang2021fast}
Chuang Zhang, Songze Chen, Zhaoli Guo, and Lei Wu.
\newblock A fast synthetic iterative scheme for the stationary phonon boltzmann transport equation.
\newblock {\em International Journal of Heat and Mass Transfer}, 174:121308, 2021.

\end{thebibliography}

\end{document}